\documentclass[a4paper,11pt]{article}
\usepackage{amsfonts}

\usepackage{graphicx}
\usepackage{amsmath}
\usepackage{amssymb}
\usepackage{latexsym}
\usepackage[colorlinks]{hyperref}
\usepackage{color}
\usepackage{float}
\usepackage{cite}
\usepackage{makeidx}
\usepackage{colortbl}
\usepackage{csquotes}
\usepackage[squaren]{SIunits}

\usepackage[textfont={footnotesize,sf},labelfont={color=blue,bf,sf},labelsep=endash]{caption}
\usepackage[position=top,labelfont={color=blue,bf,sf}]{subfig}
\usepackage{bm}

\usepackage[title]{appendix}

\newcommand{\bra}{\begin{array}}
    \newcommand{\era}{\end{array}}
\newcommand{\beq}{\begin{equation}}
\newcommand{\eeq}{\end{equation}}
\newcommand{\bqr}{\begin{eqnarray}}
\newcommand{\eqr}{\end{eqnarray}}

\def\BC{\bb C}
\def\_\BC{\bbi C}

\def\no2 {{\textstyle{n\over 2}}}

\newcommand{\si}{\sigma}

\newcommand{\pa}{\partial}

\newcommand{\lb}{\label}

\begin{document}
    \begin{titlepage}
        \setcounter{page}{1}
        \renewcommand{\thefootnote}{\fnsymbol{footnote}}

        \begin{flushright}
            %ucd-tpg:****.**\\
            %arXiv:yymm.xxxx
        \end{flushright}

        \vspace{5mm}
        \begin{center}
{\Large \bf {
Electronic Properties of Graphene Quantum Ring \\
with Wedge Disclination
%with Magnetic Field \\with  Potential
}}

\vspace{5mm}

{\bf Abdelhadi Belouad}$^a$,
{\bf Ahmed Jellal$^{a,b}$\footnote{\sf a.jellal@ucd.ac.ma}}
and {\bf Hocine Bahlouli}$^c$

\vspace{5mm}

{$^a$\em Laboratory of Theoretical Physics,  %Department of Physics,
Faculty of Sciences, Choua\"ib Doukkali University},\\
{\em PO Box 20, 24000 El Jadida, Morocco}
%{$^b$\em Saudi Center for Theoretical Physics, Dhahran, Saudi
%Arabia}

{$^{b}$\em Canadian Quantum  Research Center,
204-3002 32 Ave Vernon, \\ BC V1T 2L7,  Canada}

{$^c$\em Physics Department,  King Fahd University
of Petroleum $\&$ Minerals,\\
Dhahran 31261, Saudi Arabia}

%\vspace{30mm}

\vspace{3cm}

\begin{abstract}
We study the energy spectrum and persistent current  of
charge carriers confined in
a graphene quantum ring geometry of radius $R$ and width $w$ subjected to a magnetic flux.  We consider the case where the crystal symmetry is locally modified by replacing a hexagon by a
pentagon, square,  heptagon or octagon.
To model this type of defect we include appropriate boundary conditions for the angular coordinate. The electrons are
confined to a finite width strip in radial direction by setting infinite mass boundary conditions at the edges of the strip. 
The solutions are expressed in terms of Hankel functions and their asymptotic behavior allows to derive quantized energy levels in the presence of an energy gap.
We also investigate the persistent currents that appear in the quantum ring and how wedge disclination influences different
quantum transport quantities.

\end{abstract}
\end{center}
\vspace{5cm}

\noindent PACS numbers:   81.05.ue, 81.07.Ta, 73.22.Pr

\noindent Keywords: Graphene, quantum ring, wedge disclination, energy levels,  persistent currents.

\end{titlepage}
%\newpage
%%%%%%%%%%%%%%%%%%%%%%%%%%%%%%%%%%%%%%%%%%%%%%%%%%\zeta
\section{Introduction}

Graphene is a two-dimensional lattice of carbon atoms that
are arranged in a honeycomb pattern forming a hexagonal lattice \cite{Novoselov2004}. After its experimental discovery,  %Geim2007}
graphene has attracted a of attention, which is in part is due to its exotic electronic properties. %\cite{Castro2009}. 
Some of these include high electron mobility, %\cite{Wang2010}, 
excellent conductivity, % \cite{Novoselov2005}, 
peculiar tunneling phenomena % \cite{Katsnelson2006,Stander2009} 
and
other interesting transport and structural properties, which would be too many to list \cite{Castro2009}.
Graphene is interesting from a fundamental research perspective, as well as for potential technological applications \cite{Geim2009}, for example, in the study of liquid crystals \cite{Blake2008}, solar cells \cite{Wang2008} and high-frequency electronic devices \cite{Lin2010}.
Particles in graphene near high symmetry $K$ points are described by a low
energy effective model, the Dirac-Weyl Hamiltonian for massless charged particles. % in two-dimensions.
One of its striking features is the linear gapless energy dispersion relation represented by conic conduction and valence  bands \cite{Novoselov2004}. %\cite{Gonzalez1993,Castro2009}.

Recently, graphene-based quantum rings produced by lithographic techniques have been experimentally investigated  \cite{Russo2009}.  These systems have been studied theoretically using a tight-binding model, which does not provide simple analytical expressions for eigenstates and eigenvalues \cite{Recher2007,Bahamon2009}. It has also been shown \cite{Zarenia2009} that it is possible to  realize quantum rings of finite width by using electric fields to confine electrons in this geometry. The influence of an applied magnetic field has drawn a lot of attention \cite{Ponomarenko2008} in the context of the important and interesting physical properties of quantum rings in graphene. In this respect, one has to mention several recent theoretical studies related to different properties of quasiparticles confined in nanostructures such as
quantum dots \cite{Silvestrov2007,Schnez2008,Belouad2016} and quantum rings \cite{Petrovic2013,Costa2014}.
However, some of the most interesting effects on quasi-particles in graphene are due to the presence of a disclination defect, which can significantly alter the electronic structure, magnetic and transport properties %\cite{Gonzalez1993,
 \cite{Castro2009}.
The desire to employ graphene for studying curvature effects is motivated by the simplicity of the Hamiltonian and the important potential applications of graphene in nanoelectronics and potential future quantum computing devices \cite{Castro2009}.
An individual dislocation in free-standing graphene layers has been imaged using transmission electron microscopy \cite{100}.
Topological defects resulting from either kinetic factors or substrate imperfections have also been reported for epitaxial graphene
grown on SiC \cite{110}, Ir (111) \cite {130} and polycrystalline Ni surfaces \cite{140}.

In this paper we follow a similar set of ideas and consider a quantum ring of graphene with a wedge dislocation that can be understood
from Volterra’s cut-and-glue constructions \cite{Furtado1994} 
The calculations are performed in the continuum approximation limit in the vicinity of the Dirac points. After a general model description,
we solve the model taking into account the radial and angular degrees of freedom. We find eigenspinors in terms of the Hankel functions showing  quantized energy levels in the asymptotic limit. These results allow us to end up with a gap opening separating the conduction and valence bands.
We also find analytical expressions for  the persistent currents as a function of ring radius, total momentum, magnetic field, width of quantum ring and an integer index~$n$. Such an index is induced by disclination defect and quantifies the curvature of our geometry.
To investigate the behavior of our system, we provide numerical studies for a suitable selection of the physical parameters characterizing our system.

The manuscript is organized as follows. In section 2, we present our theoretical model based on
the Dirac Hamiltonian to describe the new geometry obtained via Volterra construction.
By introducing a magnetic flux, we give the  analytic solutions
for eigenvectors and  quantized eigenenergies in section~3. We then explicitly determine
the corresponding persistent currents generated by rotating our system in section~4.
In section 5, we discuss different numerical results  related to the energy spectrum and the persistent currents. Section 6 contains
a summary of the main results and conclusions.

%%%%%%%%%%%%%%%%%%%%%%%%%%%%%%%%%%%%%%%%%%%
\section{Theoretical model}
%%%%%%%%%%%%%%%%%%%%%%%%%%%%%%%%%%%%%%%%%%%

\begin{figure}[!hbt]
  \center
  \includegraphics[width=4.1cm]{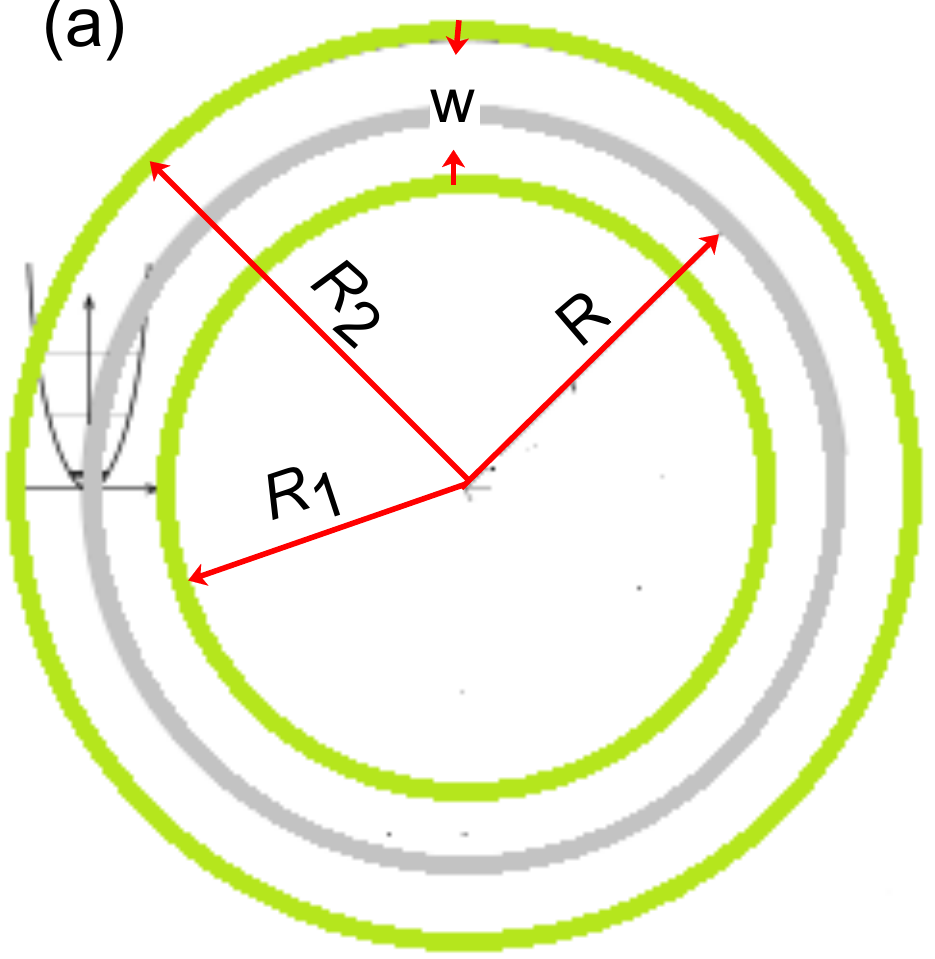} \hspace{3cm} \includegraphics[width=5.6cm]{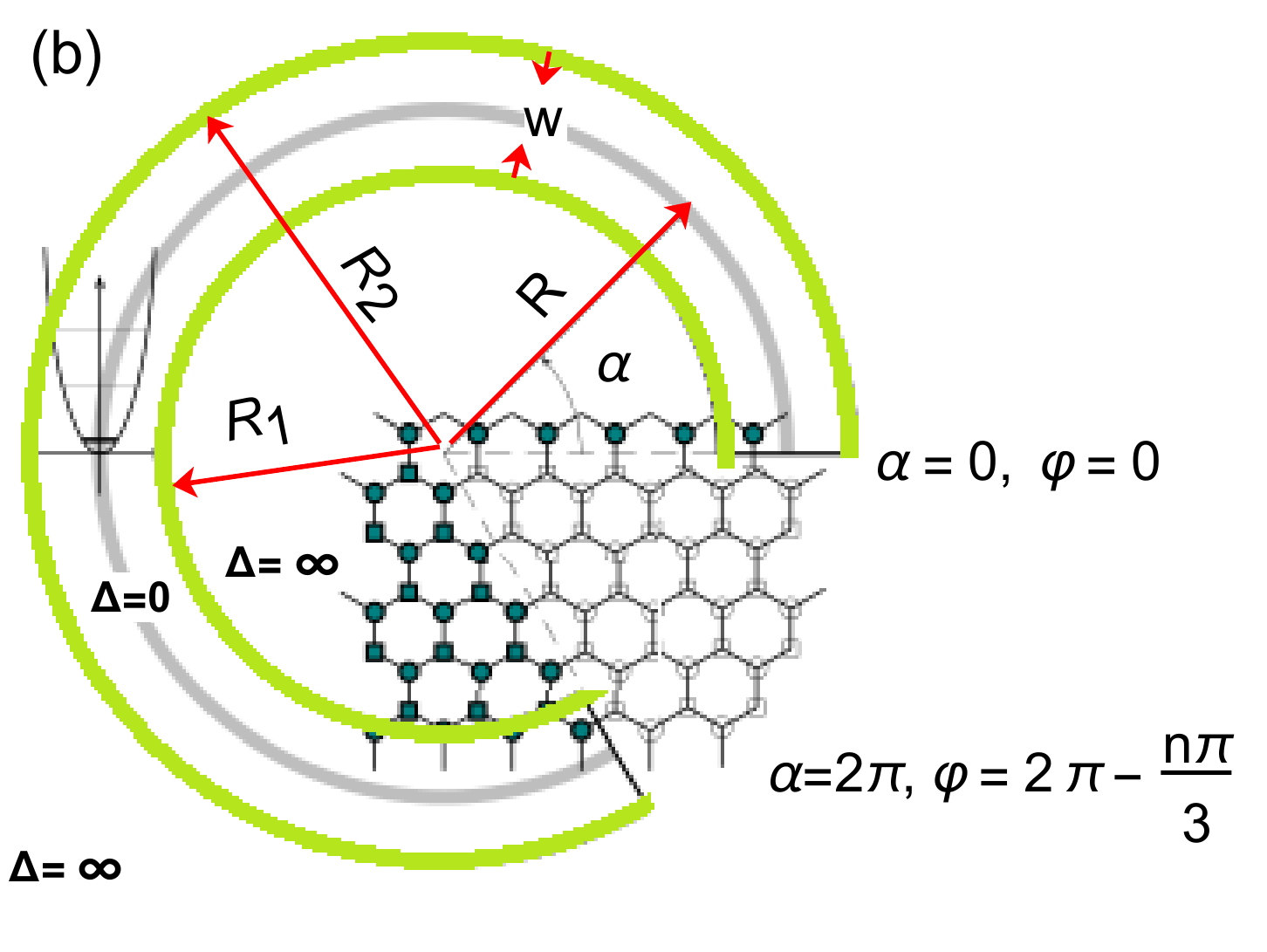}
  \caption{\sf(color online) (a): The conical ring after Volterra construction. (b):
  Unfolded plane of lattice where a wedge of angle $n\pi/3$ is removed ($n=1$ here). A potential (sketch on the left part of the ring) confines the electrons in the lowest radial mode on a ring of radius $R$, avoiding the singularity at the origin. We rescaled the angle $\varphi$ of the unfolded plane to the new angle $\alpha=\varphi/(1-\frac{n}{6})$. The carbon atoms of the removed sector are denoted by open symbols, those which remain after the cut are represented in solid symbols.
  \label{f0}}
\end{figure}

We consider a graphene quantum ring of radius $R$ and width $w$
in the presence of a magnetic flux as shown in Figure \ref{f0}a and study its electronic properties.
%Our system can be setled
%According to
 %\cite{Lin2013, Sinha2016},
We use the Volterra construction \cite{Furtado1994} to model the disclination defect of our system  by the regularized  rings of radius $R_1$ and $R_2$ around the apex and the removed wedge disclination as presented in %is removed via the Volterra construction, see
Figure \ref{f0}b.

%%%%%%%%%%%%%%%%%%

% Our starting point is the description of $\pi$-electrons
 %in monolayer graphene. Electrons

Particles in graphene have an electronic band structure with low energy band
crossings at two inequivalent high symmetry $K$ and $K'$ points,
 % \cite{Saito1998}.
at low energies one can neglect contributions away from Dirac points.
 %$K$ and $K'$.
Because of translational symmetry the Hamiltonians at the two different Dirac points can be described independently as
 %. That is at low energies in we find that the Hamiltonian for low energy graphene
 %taking both inequivalent Dirac points into account
  %is given by
\begin{eqnarray}\label{e1}
H=v_{F}[\tau_{z}\sigma_{x}p_{x}+\sigma_{y}p_{y}]+\Delta(r)\tau_{z}\sigma_{z}
\label{e1}
\end{eqnarray}
where %$\vec{\sigma}=(\sigma_{x},\sigma_{y},\sigma_{z})$, $\vec\tau=(\tau_{x},\tau_{y},\tau_{z})$
$\si_i, \tau_i$
are Pauli matrices denoting the sublattice and valley degrees of freedom, respectively, $v_{F}$ is the Fermi velocity and $\Delta(r)$ is the confining  potential of the ring as shown in Figure \ref{f0}, which is defined by
\beq\lb{delta}
\Delta(r)=\left\{%
\begin{array}{ll}
    0,  & R_1\leq r \leq R_2\\
    \infty, &  \text{otherwise}. \\
\end{array}%
\right.
\eeq
%$\Delta(r)=0$ for $R_1\leq r \leq R_2$,  otherwise $\Delta (r)=\infty$,  as shown in Figure \ref{f0}.
The Hamiltonian (\ref{e1}) acts on the two spinors
%The components of the two-dimensional wavefunction
\begin{equation}\label{e2}
\Psi^\tau(r,\varphi)=
\begin{pmatrix}
  \Psi_A(r,\varphi) \\
  \Psi_B(r,\varphi) \\
\end{pmatrix}
\end{equation}
where $\Psi_{A/B}(r,\varphi)$ is the spinor describing either of the two graphene sublattices $A$ and $B$.
It is known that deformations of the honeycomb lattice enter the continuum description via fictitious gauge fields \cite{Vozmediano2010}. As we review below, topological point defects manifest themselves by spatially well-localized fluxes of the fictitious fields \cite{Gonzalez1993}.

We mathematically discuss the new geometry obtained via Volterra construction. Indeed, for the non-rotated system, the eigenvalue equation $H\psi(0)=E\psi(0)$ implies that the new spinor
\beq
\psi(\varphi) = e^{i\varphi \si_z \tau_z/2} \psi(0)
\eeq
fulfils
$ H(R_z(\varphi))\psi(\varphi)=E\psi(\varphi)$ after rotation where $R_z(\varphi)$ is a rotation matrix. This is because  the Hamiltonian
\eqref{e1} transforms as
\beq
H(R_z(\varphi)) = e^{i\varphi \si_z \tau_z/2} H e^{-i\varphi \si_z \tau_z/2}
\eeq
which can easily be checked.
Let us remove a sector from the graphene sheet (e.g. with a pentagon replacement) and
then glue the sheet back together at the edge where we cut it as shown in Figure \ref{f0}b. Now one of the cuts can be chosen by convention to be at
angle 0 and the other at angle $-n\pi/3$ where $n$ is the curvature index that quantifies disclination defects. After gluing both of these angles become the same point and are related by a rotation
of angle $-n\pi/3$. Then, the wavefunction at this point has to be single valued and gives directly the boundary condition
\begin{eqnarray}\label{e5}
\Psi(r,\alpha=2\pi)=-e^{i2\pi[1-(n/6)]\sigma_{z}\tau_{z}/2}\Psi(r,\alpha=0)
\end{eqnarray}
where we rescale the angle $\varphi$ of the unfolded plane to the new angle $\alpha=\varphi/(1-\frac{n}{6})$,  $\alpha$ is varying from $0$ to $2\pi$.
This, however, is not the complete story and one has to look slightly beyond the continuum model to find the full story. Particularly,
let us go to momentum space and cut a wedge in the reciprocal lattice and then glue the lattice back together. We then
find that for angles $2n\pi/3$ equivalent $K$ points get glued onto each other. For angles $(2n + 1)\pi/3$, however, $K$ points get
glued onto inequivalent $K'$ points (a $60^{\circ}$ rotation connects inequivalent $K$ points). Now one can switch the $K$ and $K'$
blocks in the Hamiltonian by the unitary transformation $e^{i\pi\si_y\tau_y/2}$. Keeping track of both observations one finds the boundary
condition
\begin{equation}\label{e6}
\Psi(r,\alpha=2\pi)=-e^{i2\pi[-(n\sigma_{y}\tau_{y}/4)+(1-(n/6))\sigma_{z}\tau_{z}/2]}\Psi(r,\alpha=0).
\end{equation}
For a general index $n$ 
we introduce polar coordinates ($r, \varphi$) defined in the unfolded plane according to Figure \ref{f0}b and perform two singular transformations
\begin{equation}
\Psi(r,\alpha=2\pi)=\lambda(\varphi)\mu_{n}(\alpha)\Psi(r,\alpha=0)\end{equation}
where $\lambda(\varphi)=e^{i\varphi\sigma_{z}\tau_{z}/2}$ and $\mu_{n}(\alpha)=e^{in\alpha\si_{y}\tau_{y}/4}$.
The first one $\lambda(\varphi)$ transforms $\Psi$ to a spinor that is   expressed in the local frame ($\vec e_r, \vec e_\varphi$) with unit vectors along the radial and azimuthal directions, effectively replacing  $\partial_r$ by $\partial_r+1/(2r)$ in the Hamiltonian. The second  $\mu_{n}(\alpha)$ introduces a matrix-valued gauge field into the Hamiltonian, effectively replacing $\partial_\alpha$ by $\partial_\alpha +i\frac{n}{4}\sigma_y \tau_y $. One should note that this transformation is performed to simplify the boundary conditions \cite{Lin2013}.

\section{Magnetic flux and energy levels}

At this stage, we introduce an external source related to the magnetic field. Indeed, in the effective Hamiltonian, a magnetic flux $\Phi$ crossing
through the origin is accounted for by replacing momentum $\vec{p}$ by the conical momentum $\vec{p}+e\vec{A}$ with vector potential
\begin{eqnarray}\label{e10}
\vec{A}=\frac{\Phi}{\Omega_{n}\Phi_{0}r}\left(-\sin\varphi,\cos\varphi \right)
\end{eqnarray}
where
$\Omega_{n}=1-\frac{n}{6}$ and $\Phi_0=\frac{h}{e}$ is the magnetic flux quantum. We write the Hamiltonian  (\ref{e1}) in polar coordinate and apply the transformations through $\lambda(\varphi)$ and $\mu_n(\alpha)$, namely
\beq
\tilde H(r,\alpha)=\lambda^{\dagger}\mu_n^{\dagger}H \mu_n \lambda
\eeq
to obtain the transformed Hamiltonian
\begin{eqnarray}\label{e11}
\tilde H(r,\alpha)=\hbar v_{F}\left(k_{r}-\frac{i}{2r}\right)\tau_{z} \sigma_{x}
%\nonumber\\&&
+\hbar v_{F}\left(k_{\alpha}+\frac{\Phi}{\Omega_{n}\Phi_{0}r}+
\frac{n}{4\Omega_{n}r}\tau_{z}\right)\sigma_{y}
+\Delta\tau_{z} \sigma_{z}\quad
\label{Hamiltonian}
\end{eqnarray}%\label{e12}
we have set $k_{r}=-i{\partial \over \partial r}$ and $k_{\alpha}=-\frac{i}{r\Omega_{n}}{\partial \over \partial\alpha}$.
%Written in matrix form it is
%\begin{widetext}
%\begin{eqnarray}\label{e13}
%\tilde H(r,\alpha)=
%\begin{pmatrix}
%\tau \Delta & -i({\partial \over \partial r}+\frac{1}{2r})-\frac{1}{r\Omega_{n}}({\partial \over \partial\alpha}+i\frac{\Phi}{\Phi_{0}}+i\frac{n}{4}\tau)\\
%-i({\partial \over \partial r}+\frac{1}{2r})+\frac{1}{r\Omega_{n}}({\partial \over \partial\alpha}+i\frac{\Phi}{\Phi_{0}}+i\frac{n}{4}\tau) & -\tau \Delta
%\end{pmatrix}\label{Hamiltonianmatrix}
%\end{eqnarray}
%where $\tau=\pm 1$ correspond to the $k$ and $k'$ points
%respectively. The eigenstates of equation ($\ref{e2}$) are two component spinors which, for each valley index, are given in polar coordinates by \cite{Recher2007}
%\end{widetext}
Since $\tilde H(r,\alpha)$ %the Hamiltonian \eqref{e11}
commutes with the total angular momentum $J_z=L_z+S_z$,
then we can split the eigenspinors into angular and radial parts
\begin{eqnarray}\label{e14}
\Psi(r,\alpha)=e^{ij\alpha}
\begin{pmatrix}
\chi_{A}(r)\\
i\chi_{B}(r)
\end{pmatrix}
%\label{eigenf}
\end{eqnarray}
such that  $j=m+1/2$ are the eigenvalues of $J$ and $m\in \mathbb{Z}$.
%$m=0, \pm 1, \pm 2, \pm 3, \cdots$.
Inside the quantum ring, we use $H\Psi=E\Psi$ and show that the radial components satisfy
%from the eigenvalue equation  $H\Psi=E\Psi$ we obtain
\begin{eqnarray}\label{e15}
&&
  \left[\frac{\partial}{\partial r}+\frac{1}{2r}+\frac{1}{r\Omega_n}\left(j+\beta+\frac{n\tau}{4}\right)\right]\chi_{B}(r)=\frac{E}{\hbar v_F} \chi_{A}(r)\\
  &&
  \label{e87}
  \left[\frac{\partial}{\partial r}+\frac{1}{2r}-\frac{1}{r\Omega_n}\left(j+\beta+\frac{n\tau}{4}\right)\right]\chi_{A}(r)=-\frac{E}{\hbar v_F}\chi_{B}(r)
\end{eqnarray}
%%%%%%%%%%%%%%%%%%%
where we have defined the quantum quantity
$\nu=\frac{1}{\Omega_n}\left(m+\frac{1}{2}+\beta+\frac{n\tau}{4}\right)$ and $\beta=\frac{\Phi}{\Phi_0}$ is the dimensionless flux.
Decoupling the above equations, we arrive at the Hankel differential equation for $\chi_A(r)$
\beq\label{e17}
\left[r^2 \frac{\partial^2}{\partial r^2}+r
\frac{\partial}{\partial r}+\left(\frac{Er}{\hbar v_F}\right)^2 - \left(\nu-\frac{1}{2}\right)^2
\right]\chi_A(r)=0 \eeq
giving rise to the solution of the radial part of the eigenspinors
\begin{equation}\label{e20}
\chi_{\tau}(\rho)=a_{\tau}
\begin{pmatrix}
H_{\nu-\frac{1}{2}}^{(1)}(\rho)\\ i\text{sgn}(E)
H_{ \nu+\frac{1}{2}}^{(1)}(\rho)
\end{pmatrix}
+b_{\tau}
\begin{pmatrix}
H_{{\overline \nu}-\frac{1}{2}}^{(2)}(\rho)\\ i\text{sgn}(E)
H_{\nu+\frac{1}{2}}^{(2)}(\rho)
\end{pmatrix}
\end{equation}
where $H_{\nu}^{(1,2)}(\rho)$ are Hankel functions of the (first, second) kind,
($a_{\tau}, b_{\tau}$)  are the normalization constant and we set the variable as $\rho=\frac{Er}{\hbar v_F}$.

Now we look for the eigenvalues associated to the above eigenspinors. To this end, we first determine
the coefficients $a_\tau$ and $b_\tau$
using the boundary condition of the ring induced by $\Delta(r)$ such that  $\Delta(r)\rightarrow \infty$ outside the graphene ring. 
For this purpose, we adopt the infinite mass boundary conditions 
 \cite{Recher2007,Belouad2016}
at the two points $r_1=R-w/2$ and $r_2=R+w/2$, with $R=\frac{R_2+R_1}{2}$ and $w=R_2-R_1$.
Indeed, for
$\chi_B(\rho_1)=-i\tau\chi_A(\rho_1)$ we find
\begin{equation}\label{e21}
\frac{a_{\tau}}{b_{\tau}}=-\frac{H_{\nu-\frac{1}{2}}^{(2)}(\rho_{1})+\tau \text{sgn}(E) H_{\nu+
\frac{1}{2}}^
{(2)}(\rho_{1})}{H_{\nu-\frac{1}{2}}^{(1)}(\rho_{1})+\tau \text{sgn}(E)H_{\nu+
\frac{1}{2}}^
{(1)}(\rho_{1})}
\end{equation}
and $\chi_B(\rho_2)=i\tau\chi_A(\rho_2)$ gives
\begin{equation}\label{e22}
\frac{a_{\tau}}{b_{\tau}}=-\frac{H_{\nu-\frac{1}{2}}^{(2)}(\rho_{2})-\tau \text{sgn}(E) H_{\nu+
\frac{1}{2}}^
{(2)}(\rho_{2})}{H_{\nu-\frac{1}{2}}^{(1)}(\rho_{2})-\tau \text{sgn}(E)H_{\nu+
\frac{1}{2}}^
{(1)}(\rho_{2})}
\end{equation}
where  $R$ is the ring radius and $w$ its width, see Figure \ref{f0}. We can rearrange
\eqref{e21} and \eqref{e22} to obtain
%
%By eliminating
%$a_\tau$ and $b_\tau$, we find
the energy eigenvalue equation
$z_{(1)}=z_{(2)}$, with
%where $z_j$ is given by
%$z_1=z_
\begin{equation}\label{e23}
z_{(i)}=\frac{H_{{\nu}-\frac{1}{2}}^{(i)}(\rho_{2})-\tau \text{sgn}(E) H_{\nu+\frac{1}{2}}^{(i)}(\rho_{2})}{H_{\nu-\frac{1}{2}}^{(i)}(\rho_{1})+\tau \text{sgn}(E) H_{\nu+\frac{1}{2}}^{(i)}(\rho_{1})}, \qquad i=1,2.
\end{equation}
To obtain an analytical approximation of the energies, we use the
asymptotic form of the Hankel functions for large $\rho$, including corrections up to order $1/\rho^2$. That is
\begin{equation}\label{e25e}
H^{(1)}_\nu(\rho)=\sqrt{2\pi\rho}e^{i(\rho-\nu\pi/2-\pi/4)}(1+\delta_\nu)
\end{equation}
where
\beq
\delta_\nu=-\frac{(4\nu^2-1)(4\nu^2-9)}{128\rho^2}+i\frac{4\nu^2-1}{8\rho}+O(\rho^{-2}).
\eeq
%and  $[H^{(1)}_\nu(\rho)]^*=H^{(2)}_\nu(\rho)$.
Now using the fact that $\left[H^{(1)}_\nu(\rho)\right]^*=H^{(2)}_\nu(\rho)$
and
%To gain insight on our numerical outcomes, let us derive
%an analytic expression of the eigensolution for a ring
%of w width . We use equation (\ref{e25e}) in equality
$z_{(i)}=z_{(i)}^*$, we obtain the approximate energy levels
\begin{equation}\label{e25}
E_{mn}=E_0 \left[\frac{\tau\pi}{2}\pm \sqrt{\left(\frac{\tau\pi}{2}\right)^2+
\nu^2\left(\frac{w}{R}\right)^2}\right]
\end{equation}
where we have defined  $E_0=\frac{\hbar v_F}{w}$. Explicitly, we have
\begin{equation}\label{e255}
E_{mn}=E_0 \left[\frac{\tau\pi}{2}\pm \sqrt{\left(\frac{\tau\pi}{2}\right)^2+
\frac{1}{\Omega_n^2}\left(m+\frac{1}{2}+\beta+\frac{n\tau}{4}\right)^2\left(\frac{w}{R}\right)^2}\right]
\end{equation}
which are in fact the quantized energies and reflect the basic characteristics of our system that will be  numerical analyzed.
%to underline the basic
%features of our system.

%

%%%%%%%%%%%%%%%%%%%%%%%%%%%%%%%%%
\section{Persistent current}
%%%%%%%%%%%%%%%%%%%%%%%%%%%%%%%%%

In graphene systems the persistent current (PC) can generated as follows. We
wrap up graphene ribbon to form a tube or ring 
with a magnetic flux passing through the hole, then the helical edge state
provides a robust channel for PC. On the other hand,
the existence of PC in a normal metal ring was first proposed by Buttiker, Imry and
Landauer \cite{Buttiker1983}. Recently,
with the advent of nano-fabrication techniques, several experimental investigations have been made to
confirm the existence \cite{888,999} and the periodicity
%\cite{1010,1111,1212,1313,1414}
\cite{1111,1414}
of PC in semiconductor quantum rings. As for graphene rings, there are several studies of PC
%in a graphene ring
with various geometrical shapes (without spin-orbit interaction). For
example, %a nano-torus [13],
flat rings (similar to Corbino-disks) with various types of
boundaries \cite{Recher2007,15155,16166} or a tube with zigzag edges \cite{17177} and
folding a graphene ribbon into a ring \cite{Huang}.
In general the electronic current element between the first neighbor sites $i$ and $j$ is given by \cite{Liu2002}
\beq
 I_{ij}= \frac{4e}{\hbar}\text{\bf Im} \sum_n f(E_n)c_{in}^* H_{ij} c_{jn}
\eeq
where $H_{ij}$ is the Hamiltonian of the system,   $f(E_n)$ being the
Fermi function, $c_{in}$ are the eigenvectors corresponding to the
eigenenergy $E_n$. The total persistent current of a ring
can be calculated by the following formula
\beq
I_{pc} = -\sum_n \frac{\pa E_n}{\pa \Phi}
\eeq
with the magnetic flux $\Phi$.

Motivated by the above findings, we calculate the persistent current, carried by a given electron state, for our model of quantum ring in graphene  with a disclination.
Then, as far as our system is concerned we have the relation
\begin{equation}\label{e26}
I=-\frac{1}{\Phi_0}\sum_{m,n}\frac{\partial E_{mn}}{\partial \beta}.
\end{equation}
and after some calculation, we end up with
\begin{equation}\label{e26}
I=\mp I_0\sum_{m,n} \frac{1}{\Omega_n^2}\ \frac{m+\frac{1}{2}+\frac{\tau n}{4}+\beta}{\sqrt{\left(\frac{\tau\pi}{2}\right)^2+\left(m+\frac{1}{2}+\frac{\tau n}{4}+\beta\right)
\left(\frac{w}{R\Omega_n}\right)^2 }}
\end{equation}
with $I_0=\frac{E_0}{\Phi_0}\left(\frac{w}{R}\right)^2$. This result
will be investigated numerically for a suitable choice of physical parameters.

%%%%%%%%%%%%%%%%%%%%%%%%%%%%%%%%%%%
\section{Numerical results}
%%%%%%%%%%%%%%%%%%%%%%%%%%%%%%%%%%

We consider graphene quantum ring, with inner and outer radii  $R_1$ and $R_2$, respectively, of width $w=R_2-R_1$ and a confining potential $\Delta(r)$ for the ring defined in 
\eqref{delta}
as shown in Figure \ref{f0}. Here we assume that an applied magnetic flux $\Phi$ only passes through a disk region enclosed by the inner circle (i.e. $R_1\leq r \leq R_2$), allowing us to explore the valley  energy entirely due to the magnetic flux effect. 
We will analyse the energy levels and the persistent  currents to extract some informations about our system.

\begin{figure}[!hbt]
  \center
  \includegraphics[width=7cm]{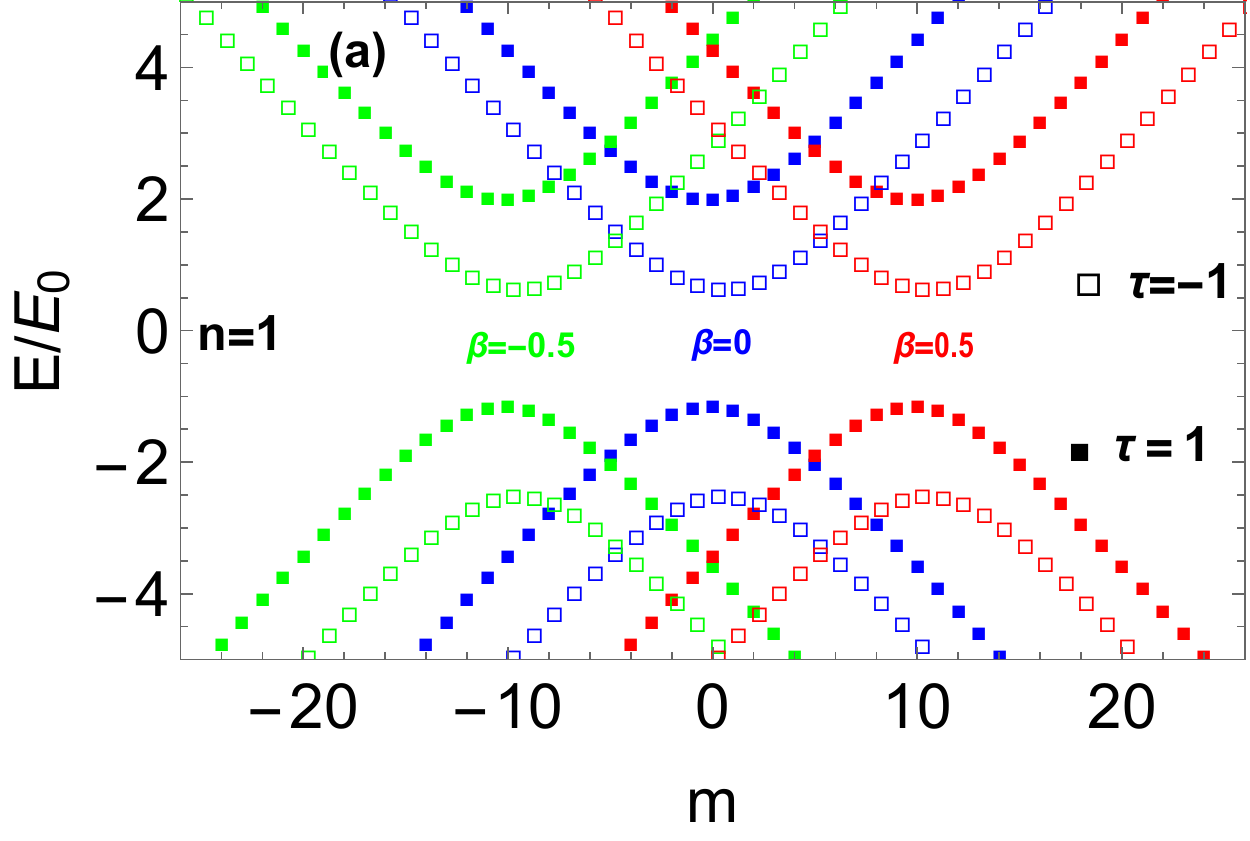}\hspace{2cm} \includegraphics[width=7cm]{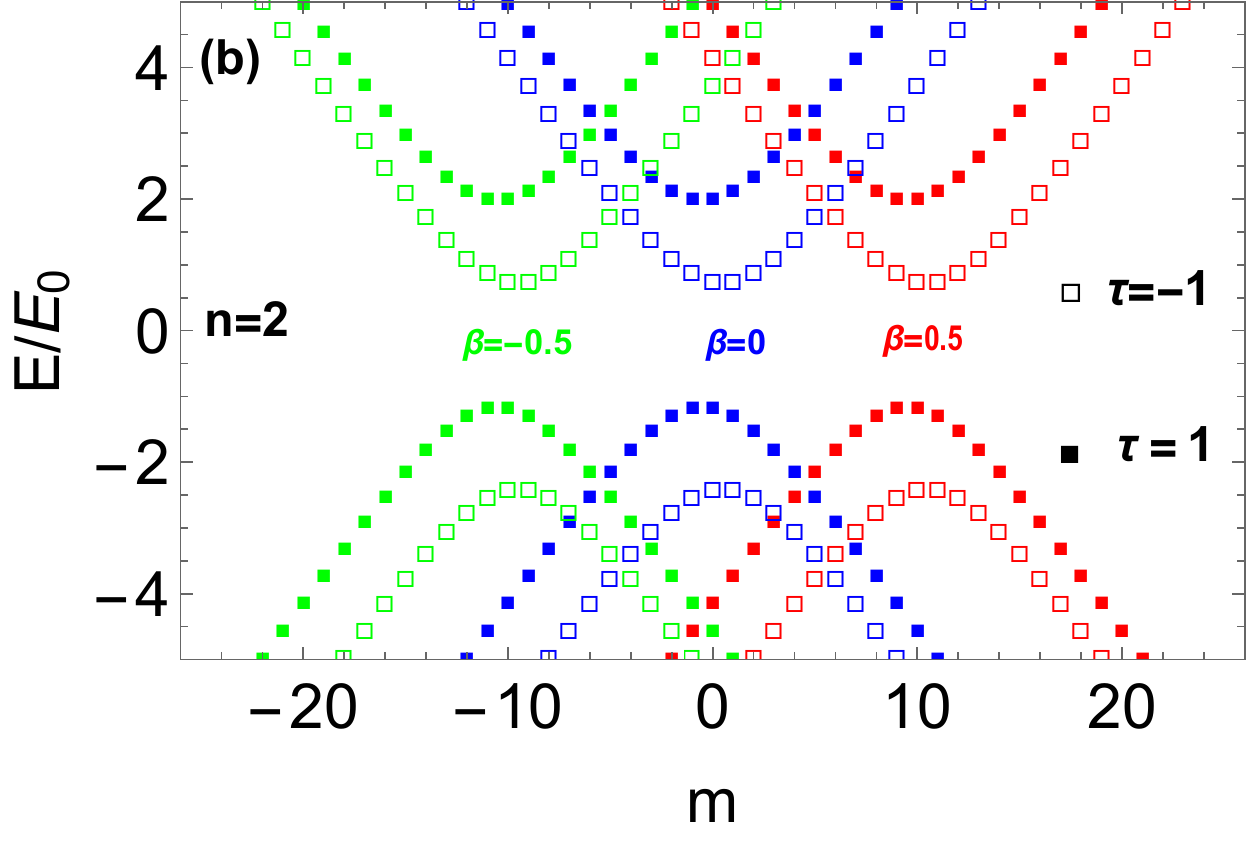}\\
  \includegraphics[width=7cm]{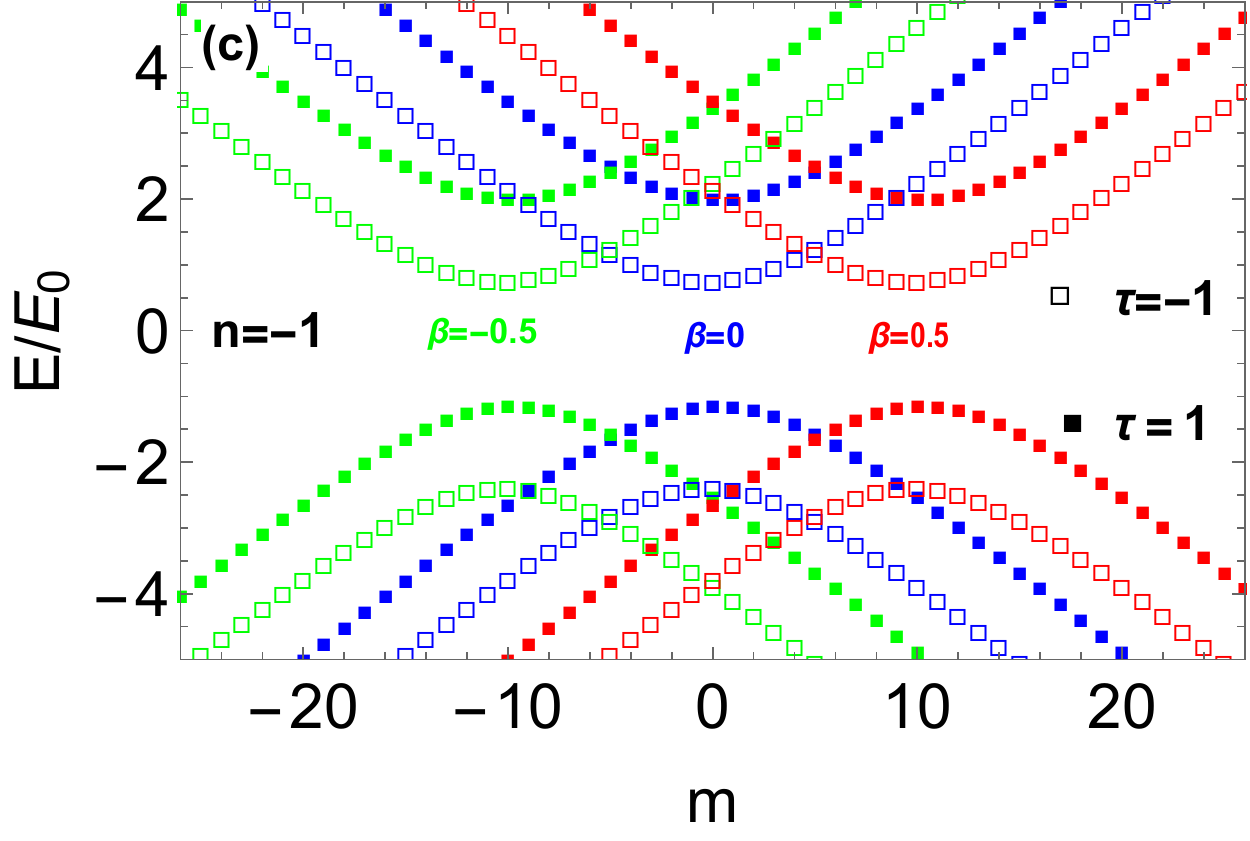}\hspace{2cm} \includegraphics[width=7cm]{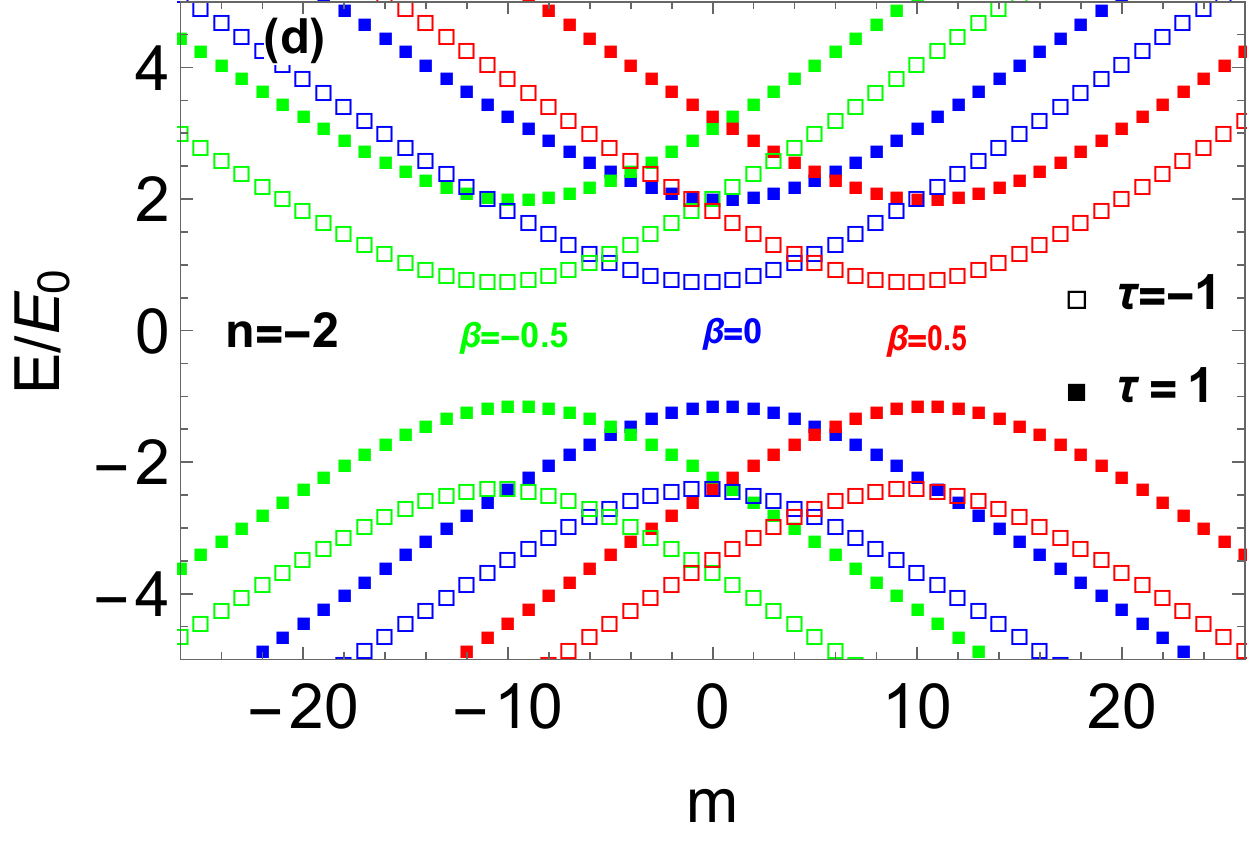}
  \caption{\sf (color online) Energy level for a graphene  with pentagon defect ($n=1$), square defect ($n=2$), heptagon defect ($n=-1$) and octagon defect (n=-2) defect as function of the quantum angular moment $m$ for different values of the magnetic flux: $\beta=0$ (blue curve), $\beta=0.5$ (red curve),  $\beta=-0.5$ (green curve).  \label{f2}}
\end{figure}

In Figure \ref{f2}, we present the results for the energy levels as function of the quantum angular moment $m$ for a graphene layer.
We show in panel (a)  the case of a pentagon defect $n=1$, in panel (b) the case of a square defect $n=2$, in panel (c) the case of a heptagon defect $n=-1$ and in panel (d) the case of a octagon defect $n=-2$. We consider  different valleys $K$ ($\tau=1$)/$K'$ ($\tau=-1$) and
three values of $\beta=-0.5$ (green curve), $\beta=0$ (blue curve) and $\beta=0.5$ (red curve). From (\ref{e25}), one may see that the spectrum exhibits an approximate minimum if the quantum angular moment takes the following form
  \beq
  m=-\left(\frac{1}{2}+\beta+\frac{n\tau}{4}\right)
  \eeq
for a given value for $\beta$ and is independent of $w$ and $R$\cite{Sinha2016}. Comparing our results with those in 
\cite{Zarenia2010}, we notice the creation of an induced pseudo-field
$\beta_1=\frac{n\tau}{4}$ due to the presence of the disclination effect. Such field  is measured by a non-zero integer index $n$ that quantifies the curvature and allows the energy spectrum to be displaced in opposite directions of the two Dirac points  $K$ and $K'$.
As a consequence, in a single valley the dependence of energy,
for both the conduction and valence bands,
as function of  $m$ is symmetric according to $\pm m$
\beq
E(\tau,m,n,\beta)= E(\tau,- m, n,-\beta).\eeq 
However, the energy  also has a
valley-index $\tau$-dependent symmetries
\beq
 E(\tau,m,n,\beta)= -E(- \tau, m,n,\beta), \qquad E(\tau,m,n,\beta)= -E(- \tau, -m,n,-\beta).
 \eeq
Furthermore, in Figures \ref{f2}(a,b,c,d) we see that there is an energy gap between
the conduction and valence bands, which is given by
 \beq
 \Delta E=2E_0\sqrt{\left(\frac{\tau\pi}{2}\right)^2+\nu^2\left(\frac{w}{R}\right)^2}
 \eeq
and depending on different values of the physical quantities
such as quantum angular moment $m$, curvature index $\beta_n=\beta+\frac{n\tau}{4}$, quantum ring parameters $w$, $R$ and valley index $\tau$.

 %%%%%%%%%%%%
 \begin{figure}[!hbt]
  \center
\includegraphics[width=7cm]{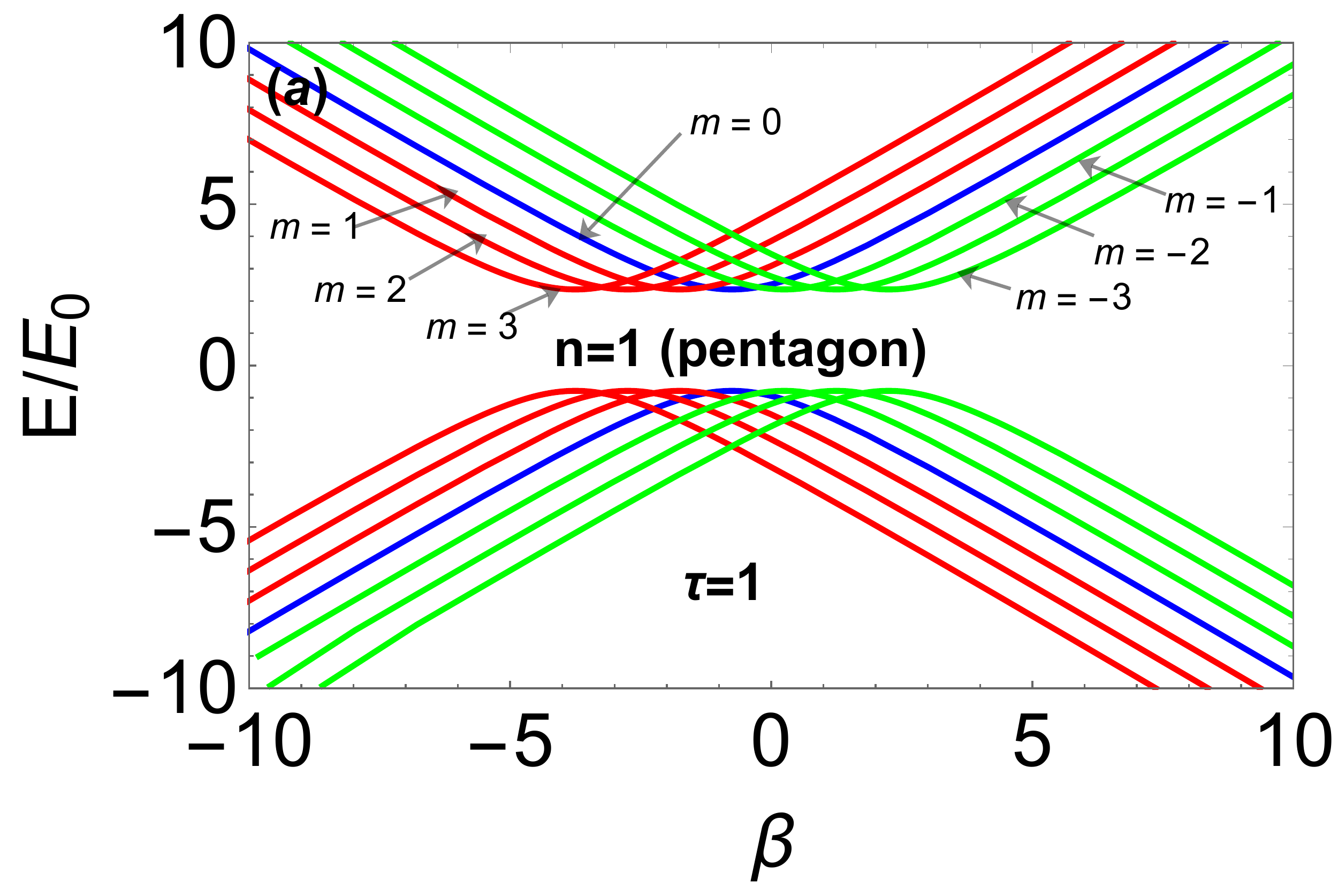}\hspace{2cm} \includegraphics[width=7cm]{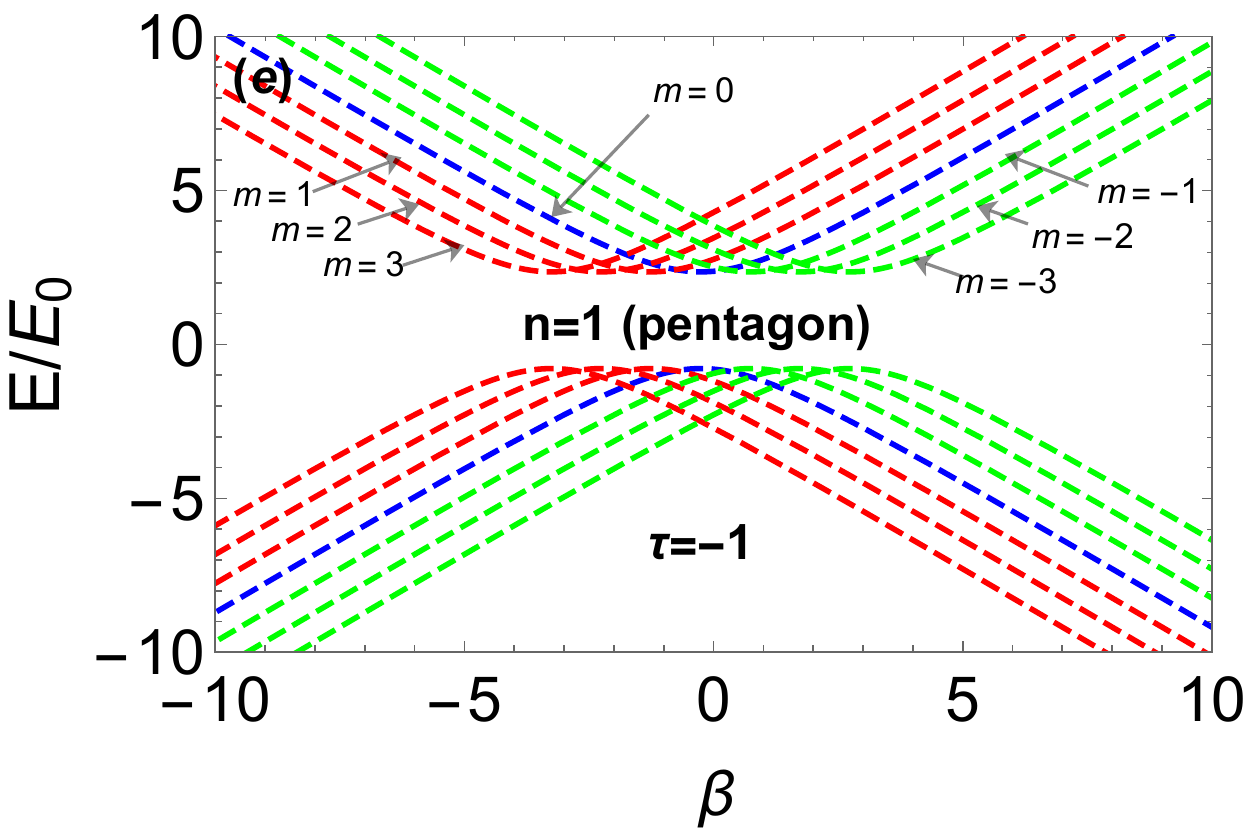}\\
\includegraphics[width=7cm]{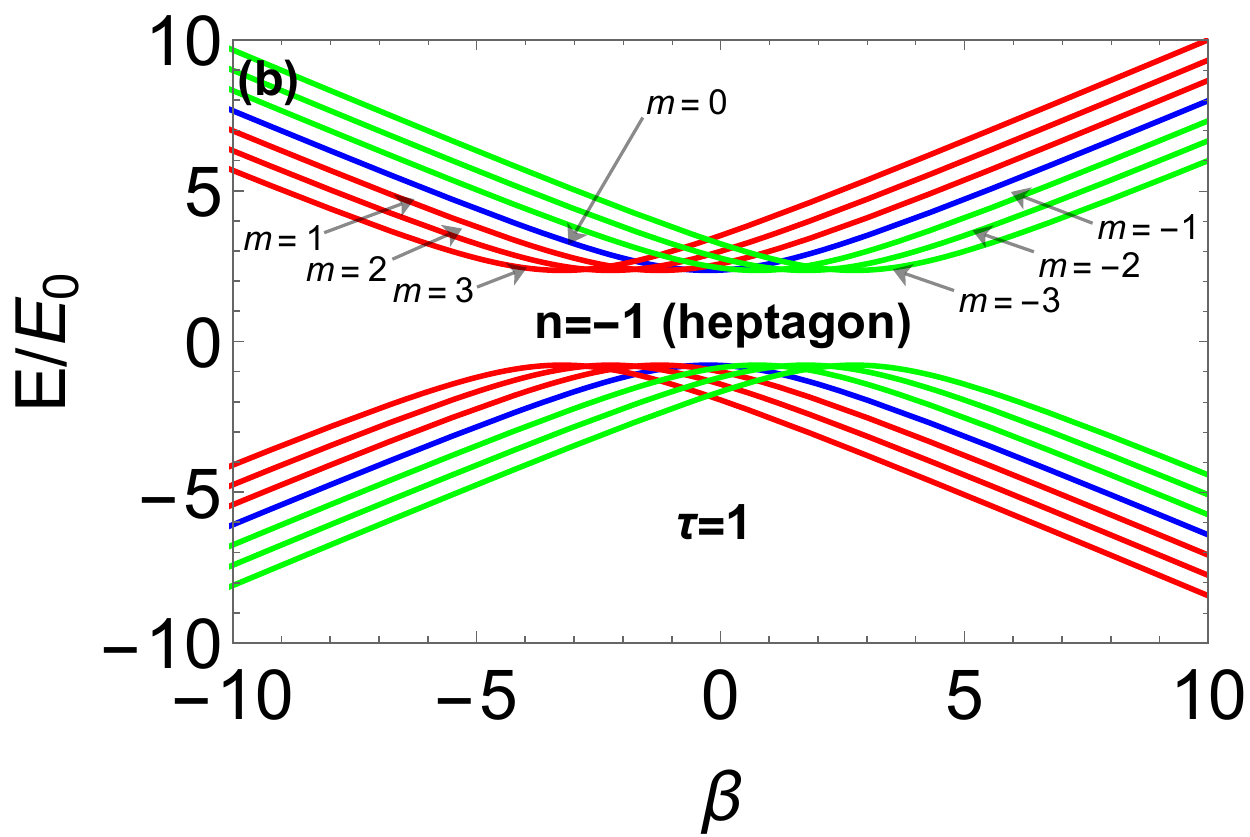}\hspace{2cm} \includegraphics[width=7cm]{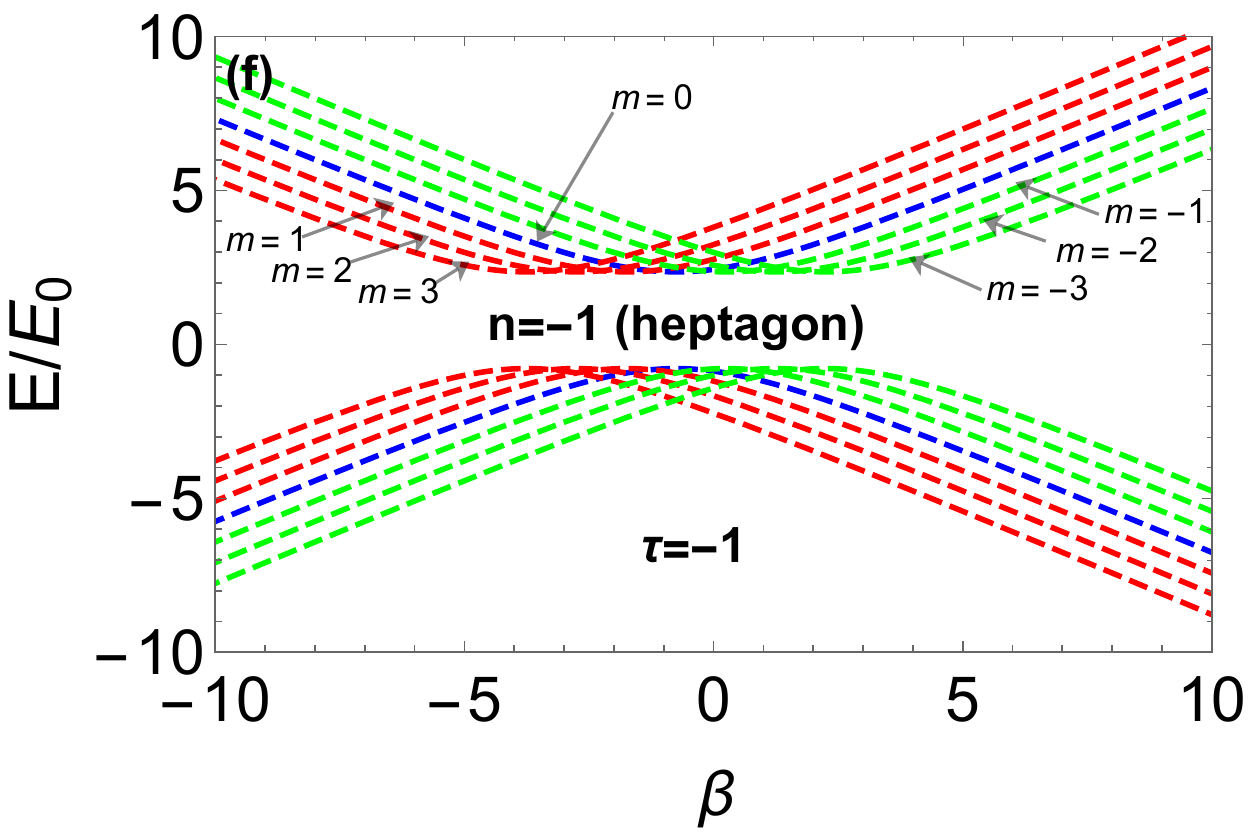}\\
\includegraphics[width=7cm]{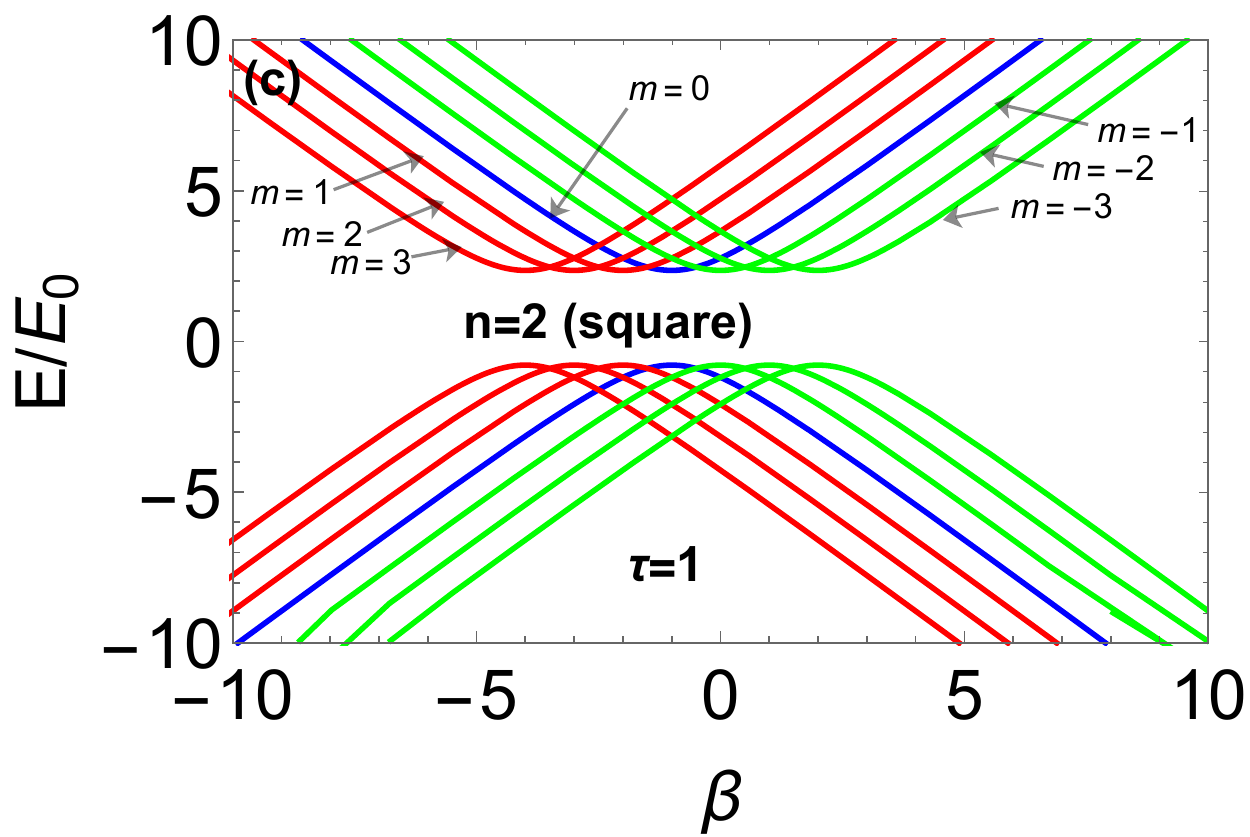}\hspace{2cm} \includegraphics[width=7cm]{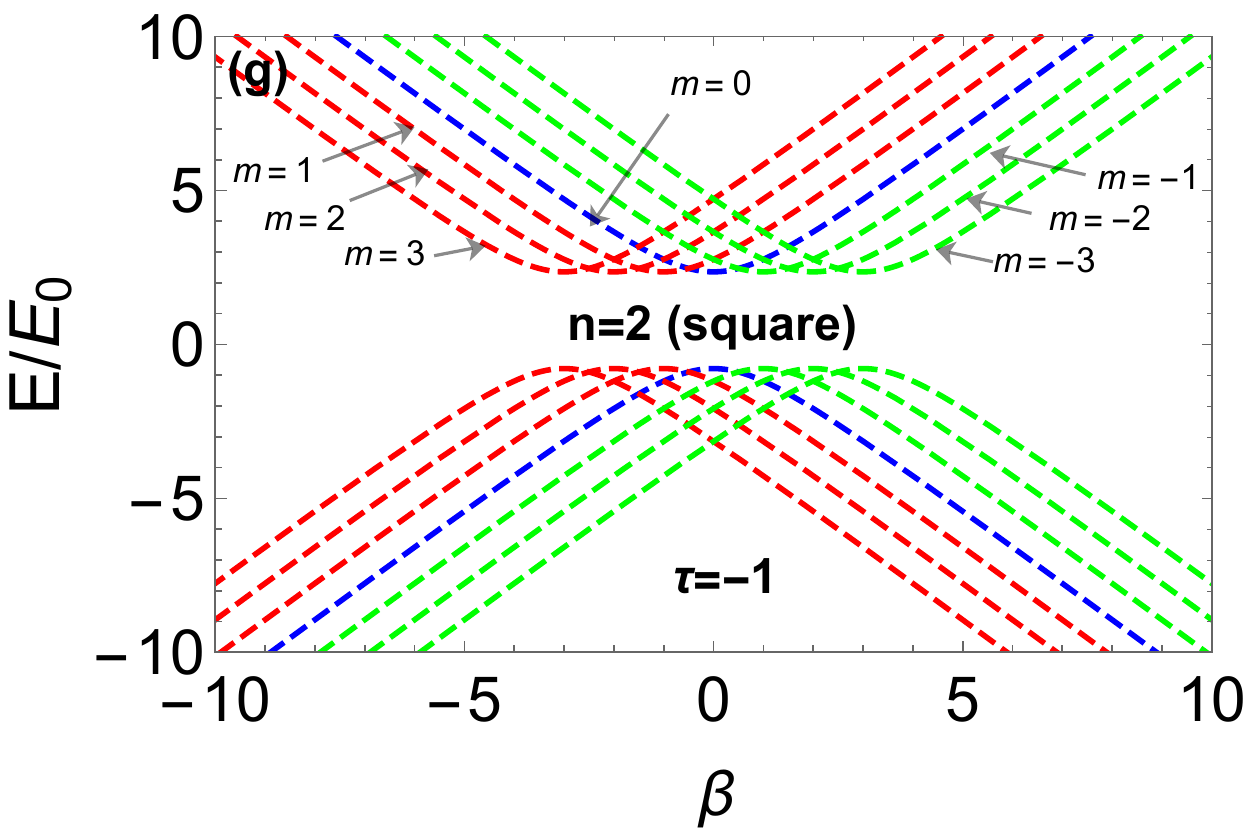}\\
\includegraphics[width=7cm]{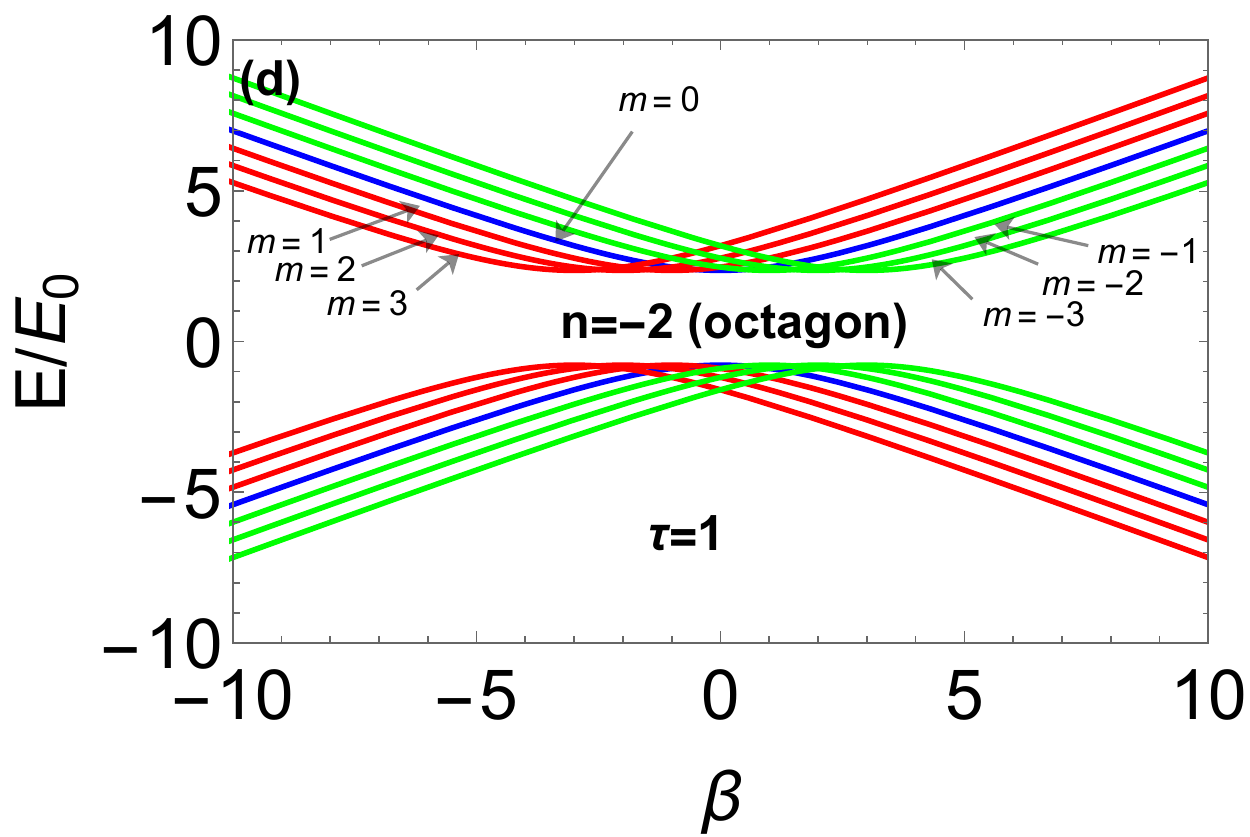}\hspace{2cm} \includegraphics[width=7cm]{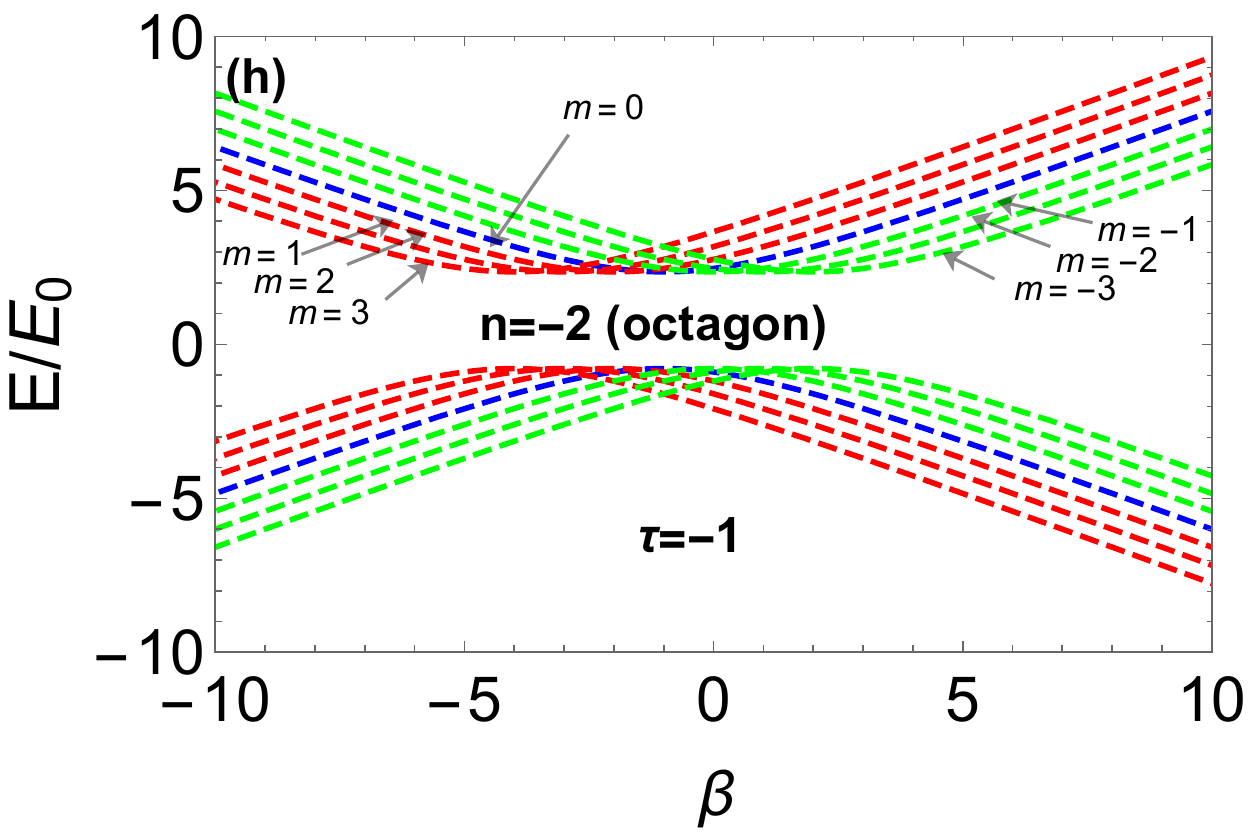}
\caption{\sf (color online) Energy levels with $m=0$ (blue), $m=1, 2, 3$ (red) and $m=-1, -2, -3$ (green) as function of magnetic flux $\beta$
%for single layer graphene quantum ring defect 
with  $R=30$ nm,  $w=5$ nm for pentagon (n=1) panels (a,e) , heptagon ($n=-1$) panels (b,f), square (n=2) panels (c,g),  octagon ($n=-2$) panels (d,h). The left panels are for $\tau=1$ and the right ones are for $\tau=-1$.\label{f3}}
\end{figure}

Figure \ref{f3} shows the energy levels as a function of the magnetic flux $\beta$ for different valleys $K$ ($\tau=1$) and $K'$ ($\tau=-1$) while the disclination is defined by the curvature index ($n=-2, -1, 0, 1 , 2$). In the absence of magnetic flux ($\beta=0$), the states with an quantum angular number $m$ in the valleys $K$ and $K'$ have different energy for the valence and conduction bands
\beq
E(\tau,m,\beta,n)=E(\tau,-m,\beta,n)
\eeq
that is because of the deformation of graphene. For $\beta\neq 0$, the states with an opposite $m$ in $K$ and $K'$  
have the same energies 
\beq
E(\tau,m,\beta,n)=E(\tau,-m,-\beta,n). 
\eeq
However, it is interesting to notice that the degeneration may still possibly remain for a non-zero flux, which is not the case for  graphene with no defect ($n=0$) \cite{Recher2007,Zarenia2010}. This comes from the periodic dependence of the energy spectrum on  magnetic flux $\beta_n=\beta+\frac{n\tau}{4}$. Physically the periodic dependence of the energy spectrum on $\beta_n$ originates from the combined effect of graphene defect and the presence of a magnetic flux, which affects the phase of the wavefunction.

\begin{figure}[!hbt]
  \center
\includegraphics[width=7cm]{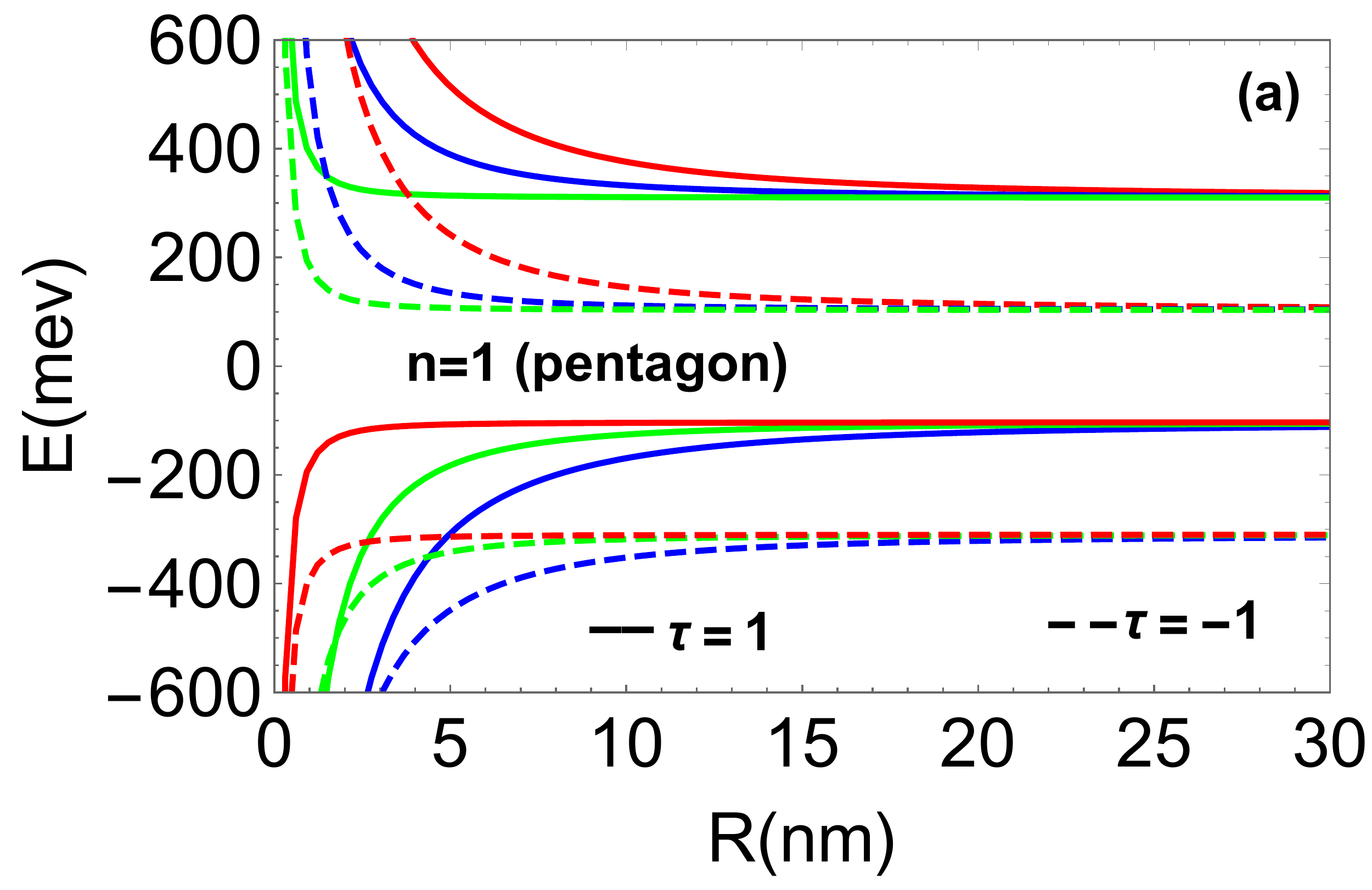}\hspace{2cm} \includegraphics[width=7cm]{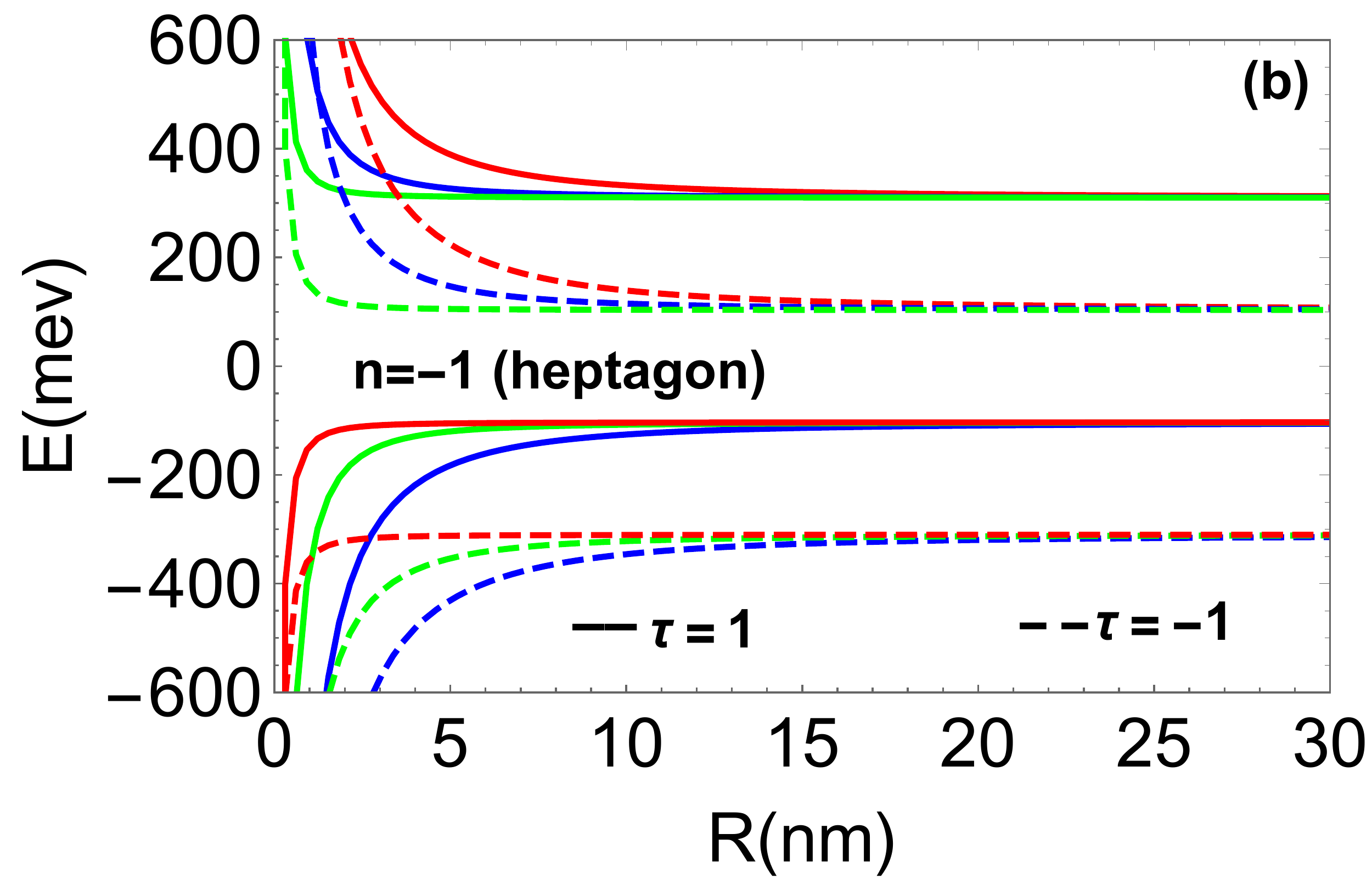}\\
\includegraphics[width=7cm]{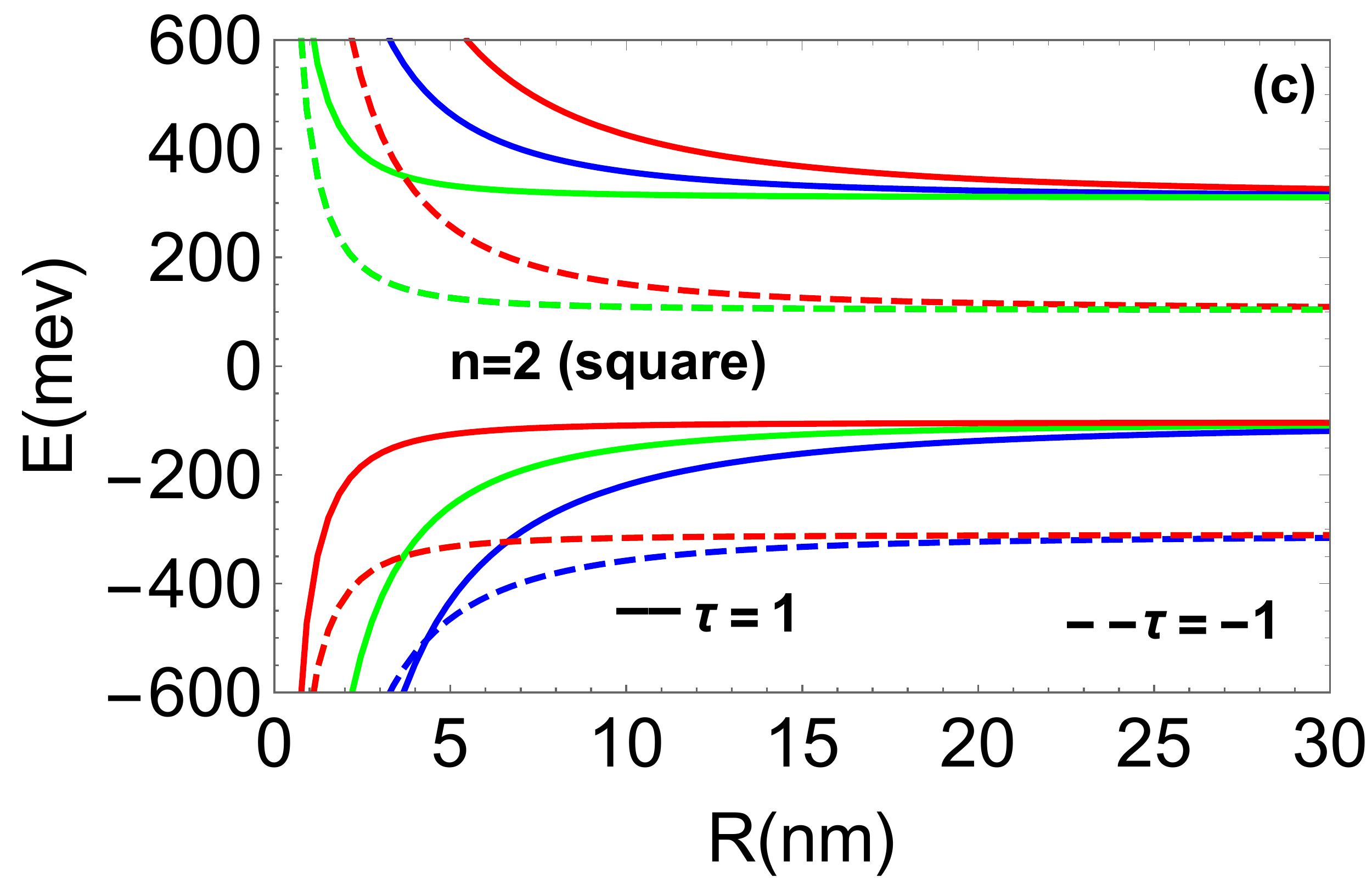}\hspace{2cm} \includegraphics[width=7cm]{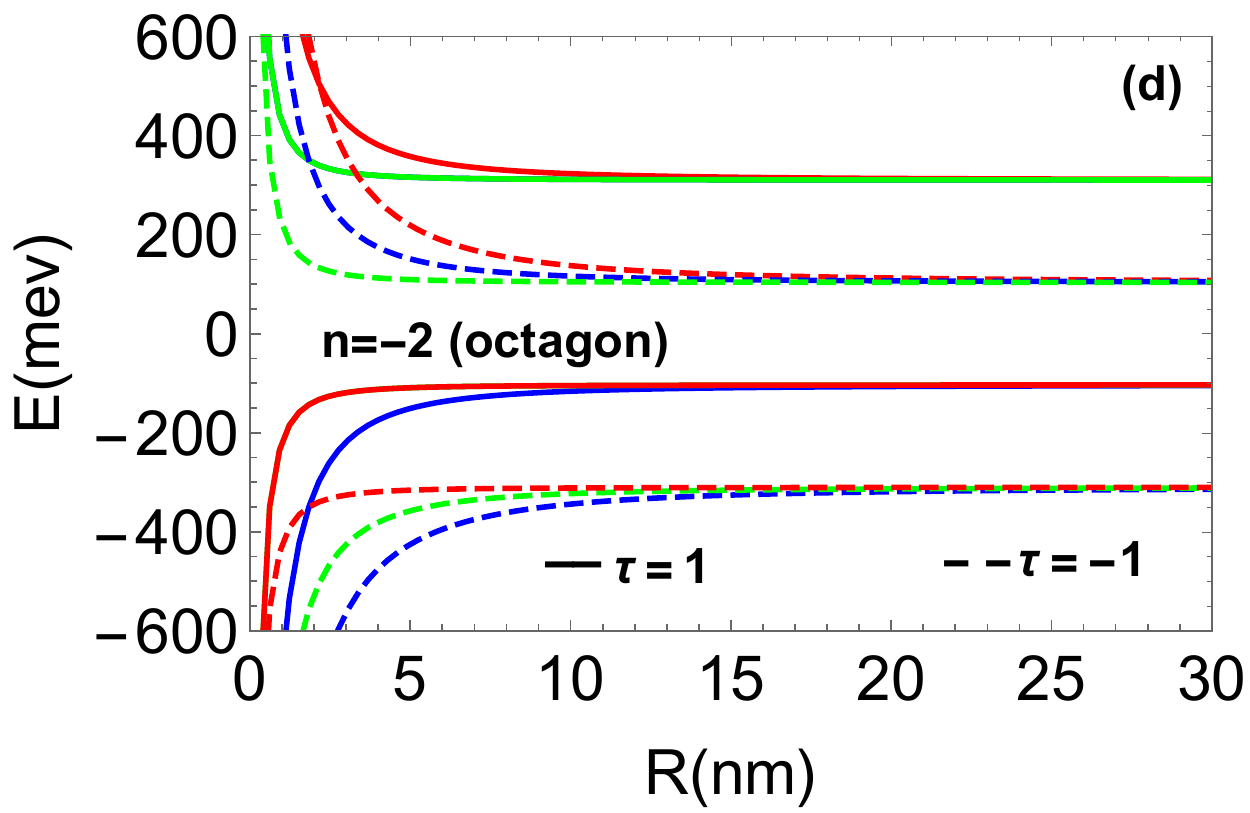}\\
\caption{\sf(color online)  Energy levels with $m=0$ (red), $m=1$ (blue) and $m=-1$ (green) %of a single layer graphene quantum ring defect as function of the ring radius R 
for $\beta=0.5$ and $w=5$ nm and different values of the index $n=1$ (pentagon), -1 (heptagon), 2 (square), -2 (octagon) for $\tau=1$ (dashed curve), $\tau=-1$ (solid curve).\label{f4}}
\end{figure}

The energy levels $E/E_0$ as function of ring radius $R$ are shown in Figure \ref{f4}
for valleys $K$ ($\tau=1$) and $K'$ ($\tau=-1$) with $\beta=1/2$, $w=5$ nm,  $m=0$ (black line), $m=1$ (blue dashed),  $m=-1$ (red line)
such that panel (a): pentagon defect ($n=1$), panel (b): square defect ($n=2$), panel (c): heptagon defect ($n=-1$), panel (d): octagon defect ($n=-2$). For small radii
 $E/E_0$
have branches converging like $\frac{1}{R}$ with an energy gap between the conduction  and  valence bands, which depends on the curvature index $n$. This behavior quantitatively differs from that 
found in standard quantum rings in graphene ($n=0$) in the absence and presence of magnetic flux \cite{Zarenia2010}. In the limit of large radii $R$,
we find that there is a constant energy gap $\Delta E= \frac{2 E_0\pi}{2}$, which does not depend on
the index $n$. As a consequence,
the symmetry 
\beq
E(\tau, m=0, n)=- E (-\tau,m=-1, n) 
\eeq
holds for all $n$.
We also find  the asymmetry $ E (\tau, n, m)\neq E (\tau,-n, m)$ is broken.

%%%%%%%%%%%%%%%%%%%%%%%%%%%%

\begin{figure}[!hbt]
  \center
  \includegraphics[width=7cm]{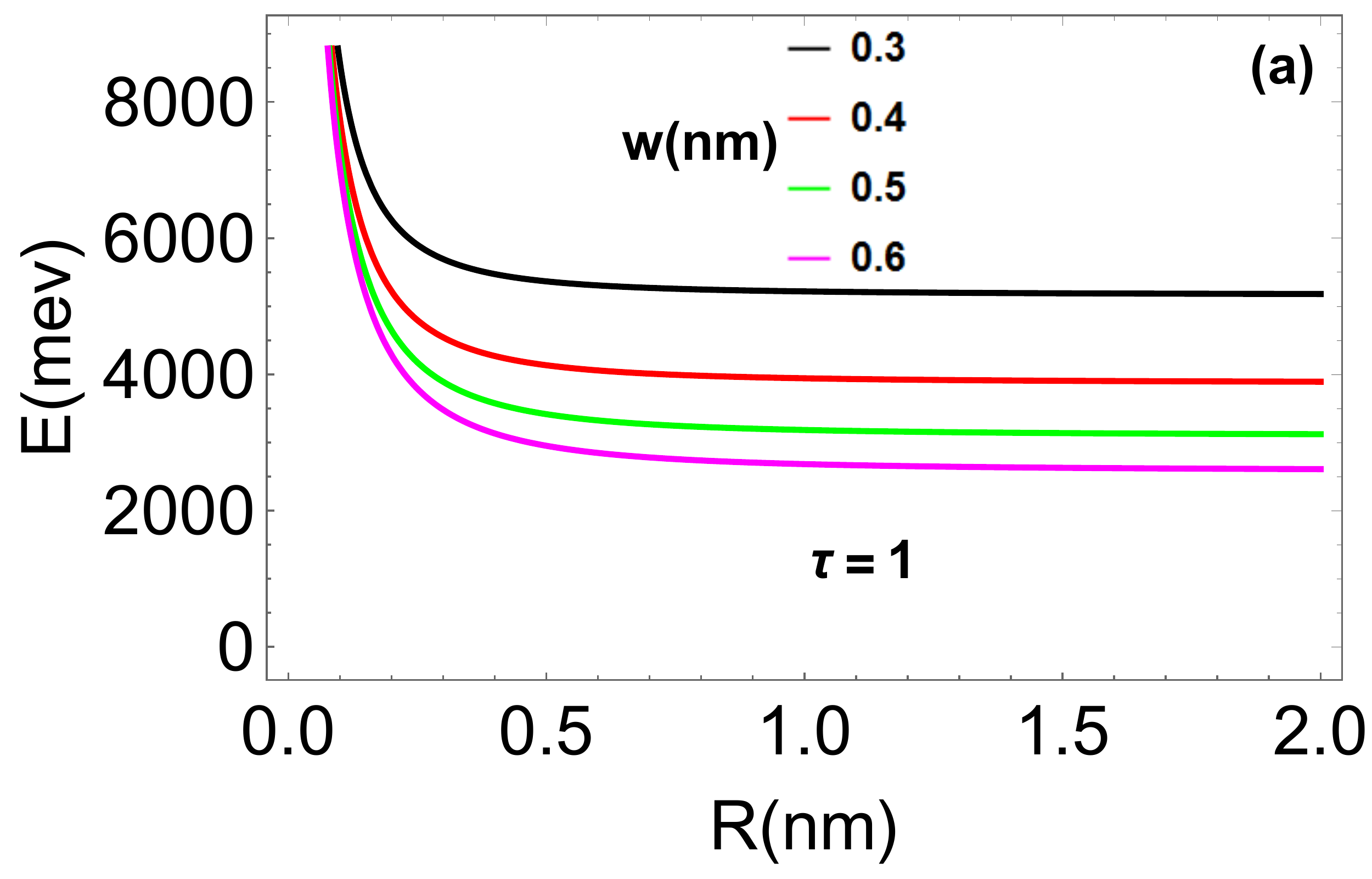}\hspace{2cm} \includegraphics[width=7cm]{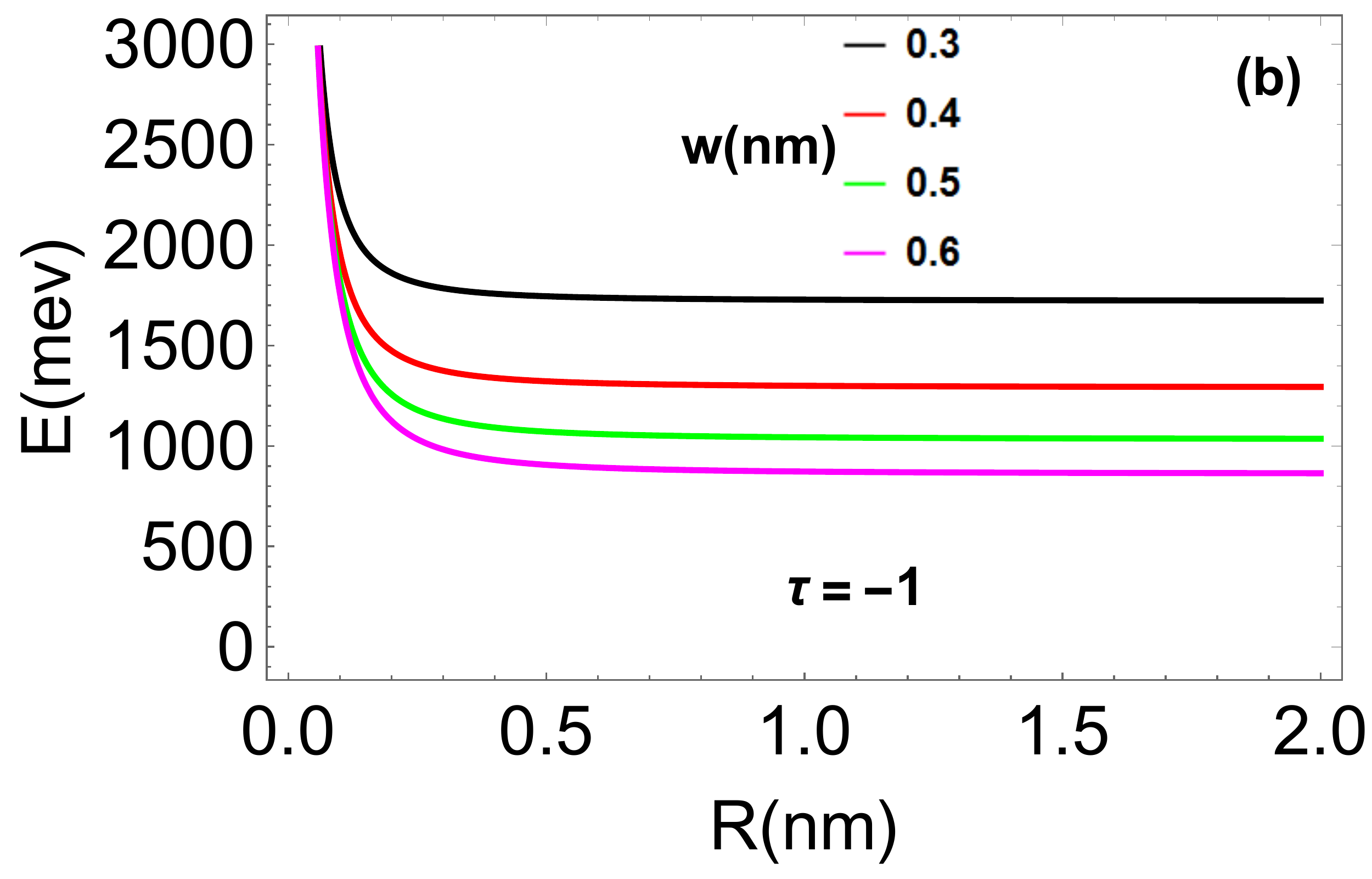}
  \caption{\sf(color online) Energy levels for a graphene layer with a square
defect as a function of $R$ for different values of width $w$ ($0.3, 0.4, 0.5, 06$) nm and $\beta=0.5$. The left panel is for $\tau=1$ and the right one is for $\tau=-1$.\label{f5}}
\end{figure}

To better understand the dependence of the energy levels $E/E_0$ on the confinement of the quantum ring of graphene defect, we present in Figure \ref{f5}
 $E/E_0$ as a function of the ring radius $R$, at constant flux $\beta=0.5$ and  $n=2$ (square) for several values of ring width $w$.
In the regime $R <0.1$ nm, $E(\tau)=E(-\tau)=0$, for $R=0.1$ nm, the energy spectrum has a maximum at $E(\tau=1)=8800$ meV for the valley $K$ and $E(\tau=-1)=3000$ meV for the valley $K'$. We find that the energy spectrum decreases with the increase in $R$.
We observe that for small values of $w$, we reach large values of $E$, such effect
is due to the strong confinement. By increasing $R$ to higher values, we notice that the energy spectrum changes only slightly and converges towards constant values. It is clearly seen that 
the energy spectrum exhibits an asymmetry $E(\tau)\neq E(-\tau)$. 

\begin{figure}[!hbt]
  \center
  \includegraphics[width=7cm]{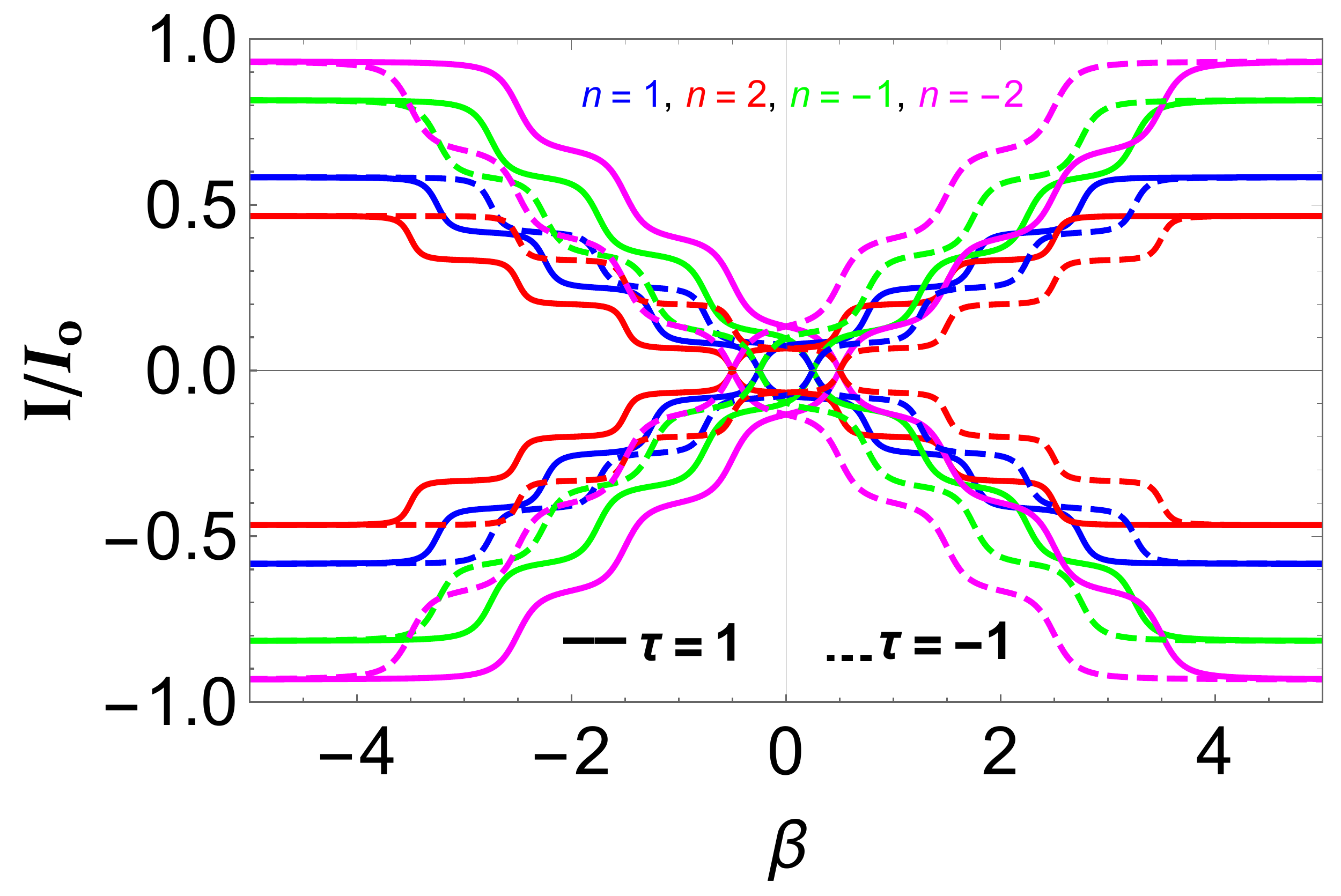}
  \caption{\sf(color online)  Persistent current $I$ as a function of magnetic flux $\beta$ for
several values of the index $n$ (-2, -1, 1, 2). The values for the other parameters are $ \frac{w}{R}=0.5$ and $\tau=\pm1$. \label{f6}}
\end{figure}

In Figure \ref{f6}, we present the persistent current $I$ as function of magnetic flux $\beta$. We consider index values $n=-2, -1, 1, 2$ for the ratio $\frac{w}{R}=0.5$. Note that solid and  dashed lines correspond to the valleys $K$ ($\tau=1$) and $K^{'}$ ($\tau=-1$). We observe there is  nonzero persistent current with zero flux for the two valleys and for any index $n$. It shows the degeneration of the valley $I(\tau)=I(-\tau)$ in the cases $\beta\leq -4$ and $\beta\geq 4$, but this degeneration is broken in the interval $-4 < \beta < 4$. Then, the persistent current oscillates as a function of magnetic flux with the same period as in the well-known Aharonov-Bohm oscillations \cite{Recher2007}.
The persistent current displays the following symmetries
\beq
I(\tau,\beta,n)=I(-\tau,-\beta,n), \qquad I(\tau,\beta,n)=-I(\tau,-\beta,n).
\eeq
Moreover, the amplitude of the current depends on the curvature index $n$.

\begin{figure}[!hbt]
  \center
  \includegraphics[width=7cm]{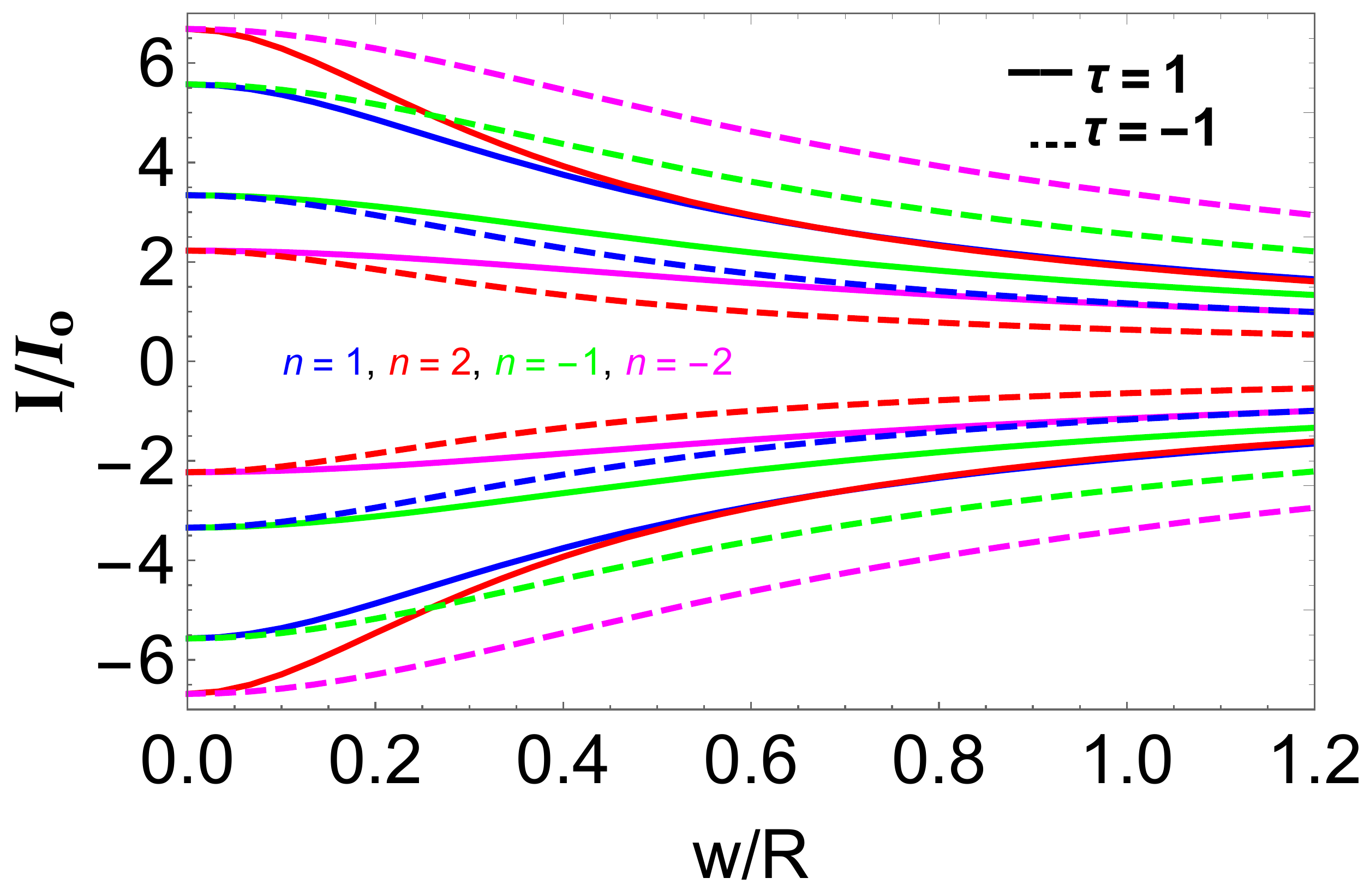}
  \caption{\sf(color online) Persistent current $I/I_0$  as a function of the ratio $w/R$ for
 valley $K$ ($\tau=1$) and $K^{'}$ ($\tau=-1$) with $\beta=0.5$. The red, green, blue and magenta curves correspond to the curvature index $n=2, 1, -1, -2$, respectively.  \label{f7}}
\end{figure}

Results for the persistent current $I$ of a graphene ring with disclination defect
as a function of ${w}/{R}$ are shows in Figure \ref{f7}. The red, green, blue and magenta curves are for the indices $n = 2, 1, -1, -2$, respectively. The solid curves and dashed curves correspond to the  valleys $K$ and $K'$, respectively. The curves have a fixed maximum $I_{max}$ for $w/R\rightarrow 0$, which can be discussed by taking into account the symmetry 
\beq
I_{max}(\tau,n)=I_{max}(-\tau,-n).
\eeq
In addition, it can be noticed that the curves are different for the same valley exhibited by our system. We can clearly see that when $w/R$ increases then the  persistent current $I$ decreases up until  $w/R \simeq 1$. The degeneracy of the valleys $K$ and $K'$ is clearly lifted. In the regime $w/R>1$, the  persistent current $I$ becomes almost constant.

%%%%%%%%%%%%%%%%%%%%%%%%%%%%%%%%%%%%%%
\section{Conclusion}
%%%%%%%%%%%%%%%%%%%%%%%%%%%%%%%%%%%%%%%

We have studied a graphene model in the shape of a quantum ring of inner radius $R_1$, outer radius $R_2$ and width $w$ subjected to a magnetic flux. %
We have considered a disclination defect using a procedure known as the Volterra process \cite{Furtado1994}. This transformation can be looked at as a cut-and-glue process where we are cut and remove a sector in the quantum ring of a graphene layer. Due to the sixfold rotational
symmetry of graphene honeycomb lattice the removed
angular sector must be a multiple of $\frac{\pi}{3}$. We have involved a topological defect in a graphene layer using a fictitious gauge field following the approach used in \cite{Vozmediano2010}
where a gauge field is introduced in the Dirac equation in order to reproduce the already known effect of the disclination on the behavior of the spinor. 
We have obtained analytical expressions for the energy levels and the corresponding eigenspinors as well as the persistent current using the infinite mass boundary condition
 \cite{Recher2007,Belouad2016}.
Our numerical results were exposed in terms variations in the ring radius $R$, total moment, magnetic field, ring width and integer-valued curvature
index $n$.

In particular, we have shown that the disclination modifies the energy spectrum and 
shifts the magnetic flux in opposite directions for the valleys $K$ and $K'$. Furthermore, the charge current is affected by the pseudo-magnetic field produced by the disclination having opposite signs at the two Dirac points. We have found an interesting behavior in the presence of a induced magnetic field, which has no analog in quantum rings with no defect ($n=0$). Indeed, for a % single layer
graphene quantum ring with a defect ($n\neq 0$), we have obtained a gap opening in the energy spectrum between the conduction and valence bands that depends on the index $n$. We also have found that the persistent current has oscillations as a function of the magnetic flux with an analogous period to the famous Aharonov-Bohm oscillations \cite{Recher2007} and obey the following symmetries  $I(\tau,\beta,n)=I(-\tau,-\beta,n)$ and $I(\tau,\beta,n)=-I(\tau,-\beta,n)$. Moreover, we have shown that the amplitude of the energy spectrum and the current depends on the index $n$ of the defects in graphene  and  on the width $w$ of the quantum ring.

\section*{Acknowledgments}
%%%%%%%%%%%%%%%%%%%%%%%%%%%%%%%%%%%%%%%%%%%%%%%%%%%%%%%%%%%%%%

The generous support provided by the Saudi Center for Theoretical
Physics (SCTP) is highly appreciated by all authors.
AJ and HB acknowledge the support of KFUPM under research group project RG181001.
HB also acknowledges discussions with Michael Vogl.

\end{document}